\documentclass[12pt]{article}
\usepackage{pdflscape}
\usepackage{tikz,natbib,xfrac,amsmath,amsthm,booktabs,multirow,subfigure,anysize,graphicx,float,enumitem,eurosym,afterpage,ragged2e,caption}
\definecolor{dark-red}{rgb}{0.4,0.15,0.15}
\definecolor{dark-blue}{rgb}{0.15,0.15,0.4}
\definecolor{medium-blue}{rgb}{0,0,0.5}
\definecolor{ChadBlue}{rgb}{.1,.1,.5}
\definecolor{ChadDarkBlue}{rgb}{.1,0,.2}
\definecolor{ChadBlue}{rgb}{.1,.1,.5}
\definecolor{ChadRoyal}{rgb}{.2,.2,.8}
\definecolor{ChadGreen}{rgb}{0,.4,0}    
\definecolor{ChadRed}{rgb}{.5,0,.5}  
\usepackage[T1,hyphens]{url}
\usepackage[colorlinks,urlcolor=medium-blue]{hyperref}
\hypersetup{
	colorlinks, linkcolor={ChadRed},
	citecolor={ChadBlue}}
\usepackage[margin=1in]{geometry}
\usepackage{xcolor}
\usepackage{bm}

\usepackage[font=small,labelfont=bf,justification=justified,singlelinecheck=false]{caption}
\usepackage{amssymb}
\usepackage{amsfonts}
\usepackage{amsmath}
\usepackage{setspace}  
\usepackage[all]{hypcap}        

\graphicspath{{figures/}}
\usepackage[english]{babel}
\usepackage{longtable}
\usepackage[singlelinecheck=off]{caption}
\usepackage[singlelinecheck=false]{caption}
\usepackage{array,makecell}
\usepackage{float}

\usepackage [autostyle, english = american]{csquotes}
\MakeOuterQuote{"}
\usepackage{chngcntr}
\usepackage{rotating, graphicx}
\usepackage{epstopdf}
\usepackage{threeparttable}
\usepackage{sectsty}
\sectionfont{\large}

\newcommand{\beginappendix}{%
	\setcounter{table}{0}
	\renewcommand{\thetable}{A\arabic{table}}%
	\setcounter{figure}{0}
	\renewcommand{\thefigure}{A\arabic{figure}}%
}
\usepackage{longtable}
\usepackage{booktabs}
\usepackage{array}
\newcolumntype{R}{>{\raggedright\arraybackslash}p{6cm}}
\newcolumntype{Q}{>{\raggedright\arraybackslash}p{6cm}}
\usepackage{tabularx}
\usepackage{dcolumn}
\usepackage[T1]{fontenc}
	\usepackage{chronosys}
	
	\makeatletter
	\newcommand{\extref}[1]{%
		\@namedef{the\@captype}{\ref{#1}}%
	}
	\makeatother
	
	\usepackage{xr}  
	\makeatletter
	\newcommand*{\addFileDependency}[1]{
		\typeout{(#1)}
		\@addtofilelist{#1}
		\IfFileExists{#1}{}{\typeout{No file #1.}}
	}\makeatother
	
	\newcommand*{\myexternaldocument}[1]{%
		\externaldocument{#1}%
		\addFileDependency{#1.tex}%
		\addFileDependency{#1.aux}%
	}
	
	\myexternaldocument{online_appendix}

\usepackage[page]{appendix} 

\title{The increasing share of low-value transactions in international trade\thanks{We thank the Department of Customs and Excise of the Spanish Tax Agency (AEAT) for providing Customs data. This research was conducted as part of the Project PID2021-122133NB-I00 financed by MCIN/AEI/10.13039/501100011033/FEDER, EU. We also gratefully acknowledge the financial support from the Basque Government Department of Education (IT1429-22). We also thank the comments and suggestions from Patricia Canto, Inmaculada Martínez-Zarzoso, and participants at Deusto Business School Brown Bag Seminar, University of Granada III Workshop on International Trends in Economic Research, and XXI Inteco Workshop.}}
	
\author{\large {Ra\'{u}l M\'{i}nguez}\thanks{M\'{i}nguez: Cámara de Comercio de Espa\~na and Universidad Antonio de Nebrija. Calle de Santa Cruz de Marcenado, 27, 28015, Madrid (Spain). Email: \href{mailto:rminguez@nebrija.es}{rminguez@nebrija.es}.  }\\ 	\and  \large {Asier Minondo}\thanks{Minondo: Corresponding author. Deusto Business School, University of Deusto, Camino de Mundaiz 50, 20012 Donostia - San Sebasti\'{a}n (Spain). Email: \href{mailto:aminondo@deusto.es}{aminondo@deusto.es}}}

\date{This version: \today \\  }

\begin{document}
\maketitle
		
\begin{abstract}
This paper documents a new feature of international trade: the increase in the share of low-value transactions in the total volume of transactions. Using Spanish data, we show that the share of low-value transactions in the total number of transactions increased from 9\% to 61\% in exports and from 14\% to 54\% in imports between 1997 and 2023. The increase in the number of low-value trade transactions is explained by the rise of e-commerce and direct-to-customer sales facilitated by online retail platforms, and the fast-fashion strategy followed by clothing firms. 
\end{abstract}

\begin{flushleft}
\end{flushleft}
		
\newpage 
\onehalfspacing
The new stage of the globalization process that began in the 1980s presents some characteristics, such as the revolution in information and communication technologies, the emergence of China as a global manufacturing exporter, and the expansion of global value chains, which were unknown in previous integration episodes \citep{krugman1995growing,rodrik2011globalization,baldwin2016great,antras2020globalisation}. This paper documents an additional novel feature of this globalization stage: the increase in the share of low-value transactions in the total number of international trade transactions.

We define a low-value transaction as the exchange of a product between a domestic and a foreign trade operator (i.e., a firm or an individual) whose value does not exceed 150 euros. This value is the de minimis threshold applied in the European Union (EU): any shipment dispatched to its recipient containing goods of a total value not exceeding 150 euros should be admitted free of import duties.\footnote{Article 27 of the Council Regulation No. 918/83 and Article 1.3 of the Council Regulation 1186/2009.} Using the universe of Spanish international trade transactions, this paper shows that the share of low-value transactions in the total number of transactions increased from 9\% to 61\% in exports and from 14\% to 54\% in imports between 1997 and 2023. 


We show that clothing and footwear dominate low-value transactions in both exports and imports, and have contributed notably to the increase in the share of low-value transactions in the total number of trade transactions between 1997 and 2023. We document that China is the most important source for low-value import transactions and the largest contributor to the increase in the number of low-value import transactions between 1997 and 2023. We also discover that there are fewer firms performing low-value transactions than large-value transactions, and the concentration of the number of transactions by firm is much higher for low-value transactions than for high-value transactions. 


Despite the large increase in the number of low-value transactions, they represent a small share of the total value of exports and imports: 0.4\% and 0.2\%, respectively, in 2023. However, if we used the less stringent de minimis threshold applied by the US, 800 USD, low-value transactions would represent a nonnegligible share of Spanish trade: 1.5\% and 0.8\% of the total value of exports and imports, and 3.8\% and 2.6\% of the value of exports and imports in consumer goods. 

We offer two explanations for the increase in the number of low-value transactions. First, the increase in the number of low-value transactions in exports until 2021 is explained by clothing and footwear firms. During the last decades, these firms have developed a so-called fast fashion strategy, characterized by the rapid adaptation of clothing displayed in the store to changing consumer preferences and the creation of scarcity with collections that end quickly. This strategy requires a frequent restocking of stores, which manifests itself in a large number of low-value export transactions between the distribution center in the home country and the stores located abroad. This frequency is higher in destinations geographically far from the distribution center, as products need more time to reach the destination. In addition, products shipped to distant destinations are more likely to be transported by air to ensure a rapid response to a change in preferences. Consistent with this argument, we show that the leading fashion retailers explained more than 60\% of the increase in the number of low-value export transactions between 1997 and 2021. Furthermore, we find that the largest number of low-value export transactions were destined to countries geographically far from Spain and products were transported by air.  

Second, the increase in the number of low-value import transactions is related to the popularization of online retail platforms, such as Shein or Temu, that allow consumers to purchase a wide range of products while providing foreign manufacturers with access to a global market. Transactions mediated through these platforms are denoted as direct to consumer, because purchases go directly from the overseas producer to the local consumer. Consistent with this explanation, we show that the increase in the number of low-value import transactions rocketed between 2021 and 2023, when the above-mentioned platforms became popular. Furthermore, these import transactions originated in China, where these online platforms were founded. The geographical distribution and administrative justification for the large number of low-value import transactions from China are also consistent with a direct-to-consumer shipment explanation. For earlier periods, we provide evidence consistent with the argument that online stores and marketplaces explain the increase in the number of low-value import transactions. Finally, we show that the increase in the number of low-value export transactions occurring from 2021 onward, which is not related to fast-fashion retailers, is also consistent with the emergence and development of online marketplaces.


Is the increase in the share of low-value transactions in the total number of trade transactions a particular feature of Spanish trade, or is it a global trend? Other studies seem to confirm the trend observed for Spain, at least for imports. \cite{fajgelbaum2024deminimis} show that the volume of de minimis imports in the United States multiplied by 10 between 2012 and 2023, and the share of de minimis imports in the total value of US imports increased from 0.002\% to 1.720\% during the same period. In the case of the EU, according to \cite{europeancommission2023deminimis}, the volume of E-commerce imports from countries outside the EU tripled between 2019 and 2022. Regarding exports, the fact that some Spanish firms are global leaders in the fashion industry could have enhanced the share of low-value transactions in the volume of export transactions, and it is reasonable to expect the share to be lower in other countries.


The increase in the share of low-value transactions in the total number of transactions has important implications for international trade. First, a growing share of imports enter the US and the EU with a value equal to or below the de minimis threshold to avoid paying tariffs is argued. This concern is particularly strong in the US since tariffs on many Chinese goods increased in 2018. \footnote{See ``Where Textile Mills Thrived, Remnants Battle for Survival'' by Jordyn Holman, The New York Times, 21 January 2024 at \url{https://www.nytimes.com/2024/01/21/business/economy/textile-mills-carolina-trade-de-minimis.html} and ``The EU targets China's Temu and Shein with proposed import duty'' by Andy Bounds Paola Tamma, Financial Times, 3 July 2024 at \url{https://www.ft.com/content/1c4c0bee-f67e-404b-877d-e0cb38faf2d6}.} In response to these concerns, the US Government announced on September 13, 2024, that it would eliminate the de minimis exception on approximately 40\% of US imports and the European Commission opened a formal proceeding against Temu on October 31, 2024.\footnote{The US Government statement is available at: \url{https://www.whitehouse.gov/briefing-room/statements-releases/2024/09/13/fact-sheet-biden-harris-administration-announces-new-actions-to-protect-american-consumers-workers-and-businesses-by-cracking-down-on-de-minimis-shipments-with-unsafe-unfairly-traded-products/}. The European Commission statemet is available at:\url{https://ec.europa.eu/commission/presscorner/detail/en/ip_24_5622}} These decisions highlight how the increase in the volume of low-value transactions could trigger important changes in trade policy.\footnote{For example, as explained by \cite{fajgelbaum2024deminimis},
between June-September 2024, Brazil, Chile, Philippines, and Turkey have either reduced the de minimis threshold, imposed tariffs on low-value transactions, or introduced new customs fees for the processing of small transactions.}

Second, the increase in the share of low-value transactions has been facilitated by the emergence of E-commerce platforms that allow consumers to choose from a wide range of products. This fact underscores the crucial role that the high-speed Internet and mobile applications play in enabling and promoting international trade. However, mobile apps may also gain access to extensive user data and governments are concerned that this puts individuals' privacy and national security at risk. Consequently, low-value transactions may emerge as a further point of contention in the ongoing trade versus security debate.

Third, E-commerce platforms allow consumers to be directly involved in international trade, and therefore their purchasing decisions have a greater weight in shaping trade flows \citep{volpe2024consumersandfirms}. If there are differences in the variables that govern consumer purchasing decisions from those of firms, it will be important to take those differences into account when designing economic policies. For example, \cite{volpe2024consumersandfirms} show that imports by individuals are not affected by trade agreements, and this paper reveals that the time sensitiveness of low-value transactions leads to a greater use of air transport. 

Fourth, low-income consumers import more low-value products than high-income ones \citep{fajgelbaum2024deminimis}. This underscores the need to consider the impact of trade policies, such as the elimination of the de minimis threshold, on income inequality. 

Fifth, many low-value goods are shipped directly from manufacturers to consumers. These shipments are transported by air and save warehouse costs, introducing a major change in the logistics of international trade. Finally, the use of air transport suggests that low-value transactions may have a greater environmental impact than the average firm-to-firm transaction \citep{biancolin2024environmentalimpact}. In addition, the environmental impact of low-value transactions could be aggravated by the fact that many E-commerce purchases are returned \citep{zhang2022returns}.  

To the best of our knowledge, this paper is the first to document the increase in the share of low-value transactions in the total number of international trade transactions. This discovery adds to the literature that has analyzed the features of the globalization stage that began in the 1980s \citep{krugman1995growing,rodrik2011globalization,baldwin2016great,antras2020globalisation}. We also add to the literature on the role of the de minimis threshold in international trade. \cite{fajgelbaum2024deminimis} document the increase in de minimis imports in the US and show that less affluent consumers are more likely to be recipients of low-value imports. We contribute to this literature by analyzing both low-value export and import transactions, quantifying the share in value and volume of low-value transactions in total trade transactions, exploring a longer time period, and highlighting some stylized facts about low-value transactions. We also offer a new explanation for the increase in the number of low-value transactions: the fast-fashion strategy followed by fashion retailers.

Our paper also speaks to the literature on E-commerce, online platforms, and direct-to-consumer shipments. Online platforms reduce barriers to finding and evaluating potential foreign consumers or suppliers, and facilitate logistics and international payments \citep{lendle2016eBay,chen2021alibaba,carballo2022onlinebusinessplatforms}. We contribute to this literature by providing evidence consistent with the argument that corporate online stores and online marketplaces contributed to the rise in the number of low-value trade transactions.


We also contribute to the literature exploring the relationship between fast fashion and international trade. \cite{fernandes2020fastfashion} show that Portuguese clothing manufacturers responded to Chinese competition by increasing the frequency of their shipments. We show that the high frequency of shipments by fashion retailers played a very important role in the increase in the number of low-value export transactions. Furthermore, we show that these transactions were concentrated in countries that were geographically far from Spain and products that were transported by air. These latter findings also link our paper to the literature that analyzes how the time sensitivity of demand determines the choice of transport means \citep{hummels2013timeastradebarrier}. The intensive use of air shipping by online retail platforms for a very wide range of products shown in our paper suggests that the choice of mode of transport can also be determined by a strategy to counteract the immediacy advantage of local offline and online stores. Finally, our paper is also related to the lumpiness of trade literature \citep{alessandria2010inventories,kropf2014fixedcostspershipment,hornok2015administrativebarrierstotrade,hornok2015pershipmentcosts,bekes2017shipment}. These papers argue that there is a trade-off between per-shipment trade costs and shipping frequency. The staggering increase in the volume of low-value transactions uncovered by this paper suggests that some firms have been able to minimize the per-shipment trade costs.

\section{Data and stylized facts}
\label{sec:data_and_stylized facts}

We obtain trade data on goods for the period 1997-2023 from the Customs and Excise Department of the Spanish Tax Agency (AEAT-Customs). Each observation captures the exchange of a product between a Spanish trade operator (a firm or an individual) and a foreign trade operator. It provides information on the direction of the trade flow (exports or imports), the Spanish trade operator's customs identification code, the product in the 8-digit Combined Nomenclature (CN) classification, the value (in euros), the country of destination or origin, the Spanish province of destination or origin, the month in which the transaction occurred, and the date the transaction was documented in customs. It does not provide an identification of the foreign partner. 

The Spanish trade data are suitable for analyzing the evolution of low-value transactions because, in contrast to other countries' databases, it does not apply a minimum value threshold to include a transaction into the database. We remove all transactions that do not involve the transfer of property in exchange for financial compensation, such as gifts, donations, or return of goods.


We define a transaction as the exchange recorded by a single observation of our dataset: the commercial exchange of a product between a Spanish and a foreign trade operator. It is important to note that our dataset captures transactions and not shipments. Therefore, an import transaction of 150 euros or less from a non-EU country recorded in our database cannot be automatically equated with a de minimis shipment since this transaction could belong to a shipment that includes multiple products.

The AEAT-Customs database combines information from two sources: Intrastat and Extrastat. Intrastat captures firms' trade flows with EU partners. In 2023, firms had to present a monthly Intrastat declaration if they accumulated an export (import) value with EU members of 400,000 euros in the current year or in the previous year. Extrastat captures firms' trade flows with non-EU countries. This information is based on the customs declaration EU-based firms must present every time they perform an export or import operation with a non-EU country. Firms must present a customs declaration regardless of the amount of trade they accumulated in the previous year or have accumulated in the current year. Due to Intrastat's data presentation threshold, AEAT-Customs does not capture all the small- and large-value transactions occurring between Spanish firms and their EU partners. This should not affect our estimations, because the below-the-Intrastat threshold trade only represented 1.8\% of total trade.\footnote{This percentage, estimated by AEAT-Customs, corresponds to 2022 (the last year with definitive data). It is available at: \url{https://sede.agenciatributaria.gob.es/Sede/estadisticas/estadisticas-comercio-exterior/principales-resultados-serie/2023.html}.} In any case, the Online Appendix shows that there are no major differences in the evolution of the number of low-value trade transactions between Spain and EU countries and Spain and non-EU countries. 

\begin{figure}[p!]
\begin{center}
\caption{Low-value transactions. Stylized facts I}
\label{fig:stylized_facts_1}
\hspace*{-2cm}
\begin{tabular}{c c}
\multicolumn{2}{c}{\small A. Number of low-value transactions and their share in the total number of transactions}\\
\small Exports& \small Imports\\
\includegraphics[height=2.2in]{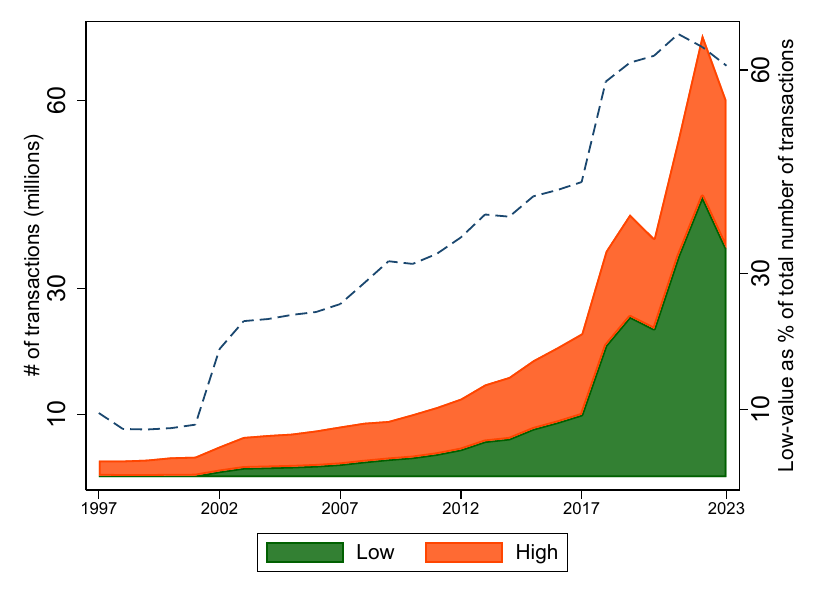}&\hspace*{-.26in}
\includegraphics[height=2.2in]{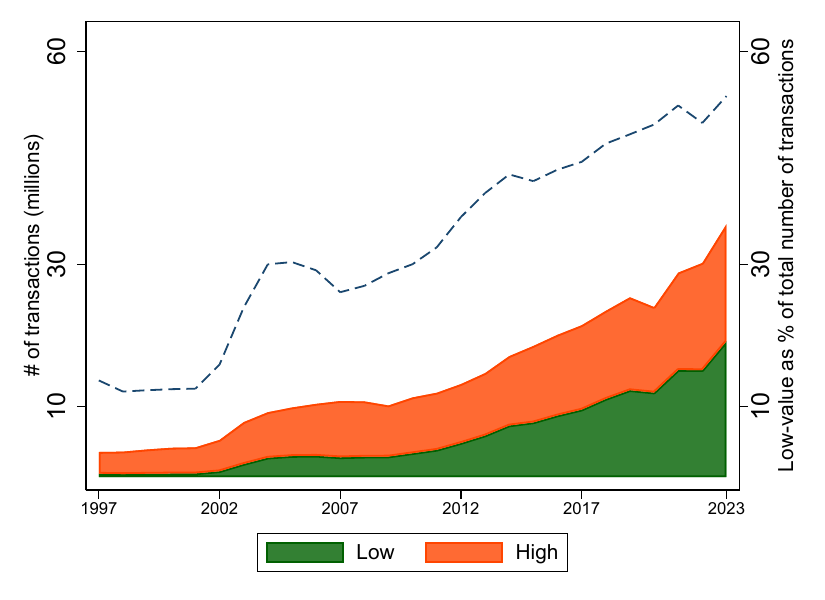}\\
\\[1em]	
\multicolumn{2}{c}{\small B. Share of low-value transactions in total trade value}\\
\small Exports& \small Imports\\
\centering
\includegraphics[height=2.2in]{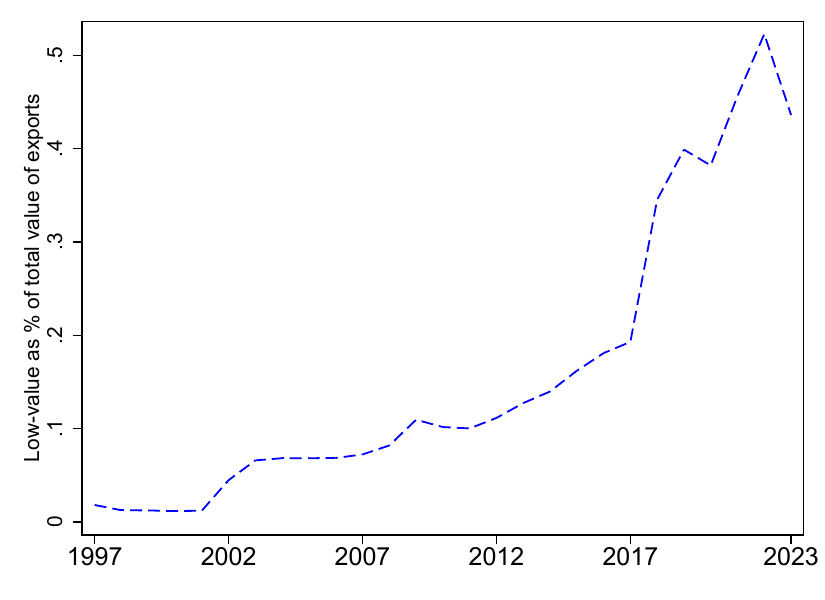}&\hspace*{-.26in}\includegraphics[height=2.2in]{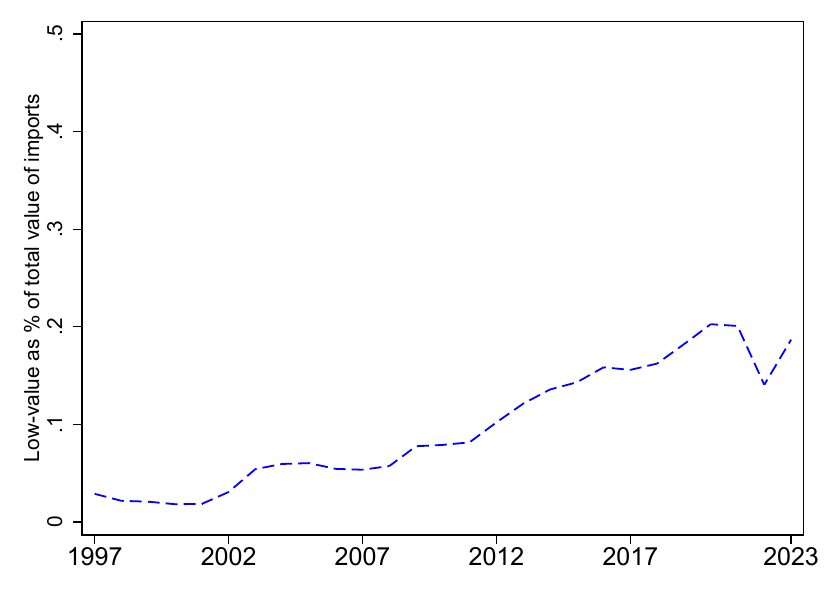}\\
\\[1em]	
\multicolumn{2}{c}{\small C. Top 20 HS 2-digit chapters by share in total number of low-value transactions (2023)}\\
\small Exports& \small Imports\\
\centering
\includegraphics[scale=.5]{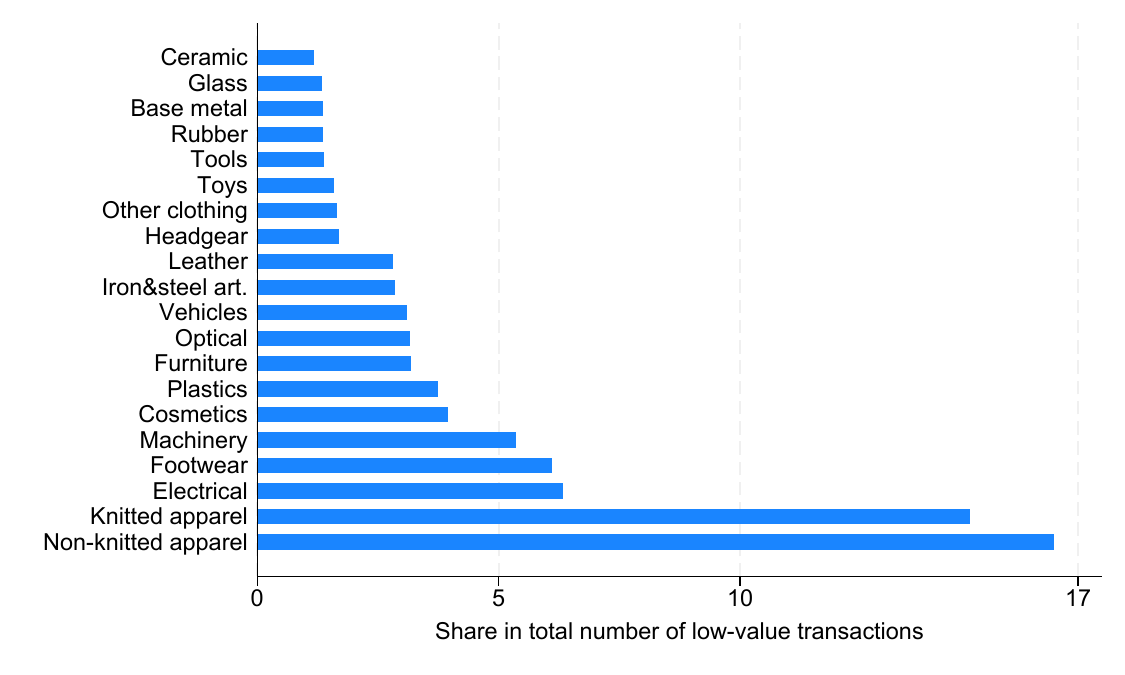}&\hspace*{-.26in}\includegraphics[scale=.5]{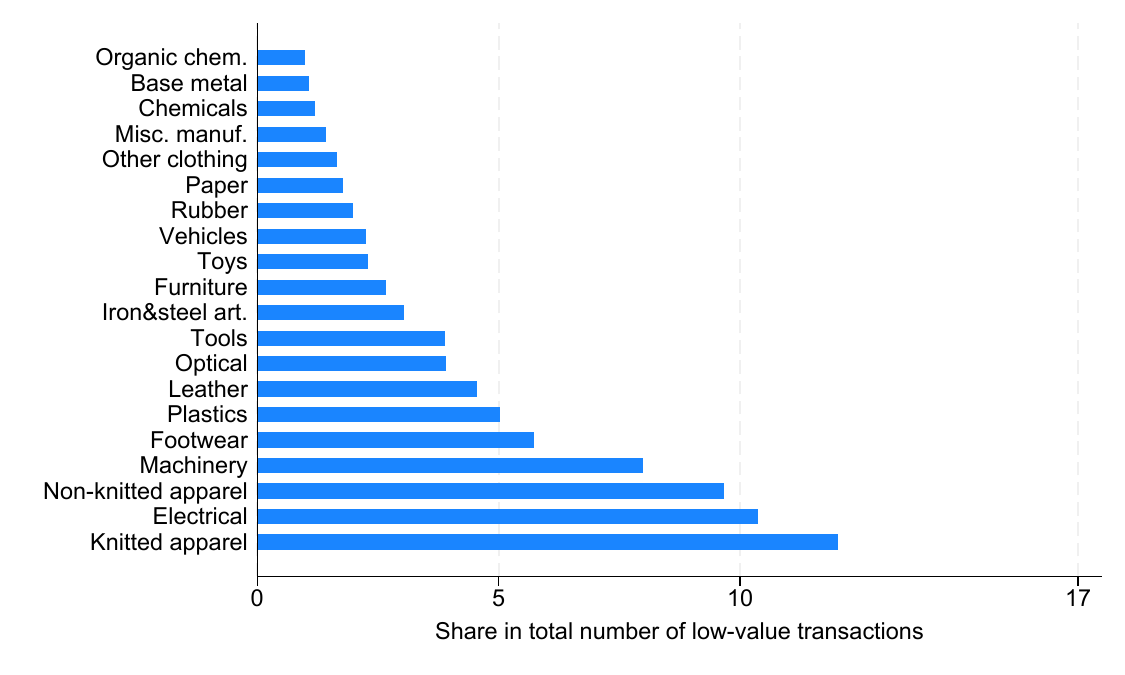}\\
\end{tabular}
\end{center}
\footnotesize Note: authors' calculations based on AEAT-Customs.
\end{figure}

Panels~A and~B of Figure~\ref{fig:stylized_facts_1} show the evolution of the number of low- (green area) and high-value (orange area) export and import transactions from 1997 to 2023, respectively. They also plot the share of low-value transactions (blue dashed line) in the total number of export or import transactions. There is a large increase in the number of transactions between 1997 and 2023: export transactions rise from 3 million in 1997 to 60 million in 2023 (they multiply by 20), while import transactions rise from 4 million in 1997 to 36 million in 2023 (they multiply by 9). 



The first stylized fact, whose documentation constitutes the main contribution of this paper, is the large increase in the share of low-value transactions in the total number of international trade transactions. It rises from 9\% to 61\% in exports and from 14\% to 54\% in imports. That is, in 2023, low-value transactions represented more than half of the total number of trade transactions.\footnote{The median values of an export and import transaction in 1997 were 4,678 and 2,776 euros, respectively. In 2023, they decreased to 77 and 114 euros.} As shown in Figure~\ref{fig:appendix_rob_deminimis} in the Online Appendix, this stylized fact is robust to using a low-value threshold adjusted for inflation, a low-value threshold based on the 800 USD de minimis value applied by the United States, and to splitting the sample into EU and non-EU members.\footnote{The exception is that the share of low-value export transactions in the total number of transactions between 2021 and 2023 decreased in non-EU27 countries and increased in EU27 countries.} 


The total value of low-value exports increased from 16 million euros in 1997 to 1,478 million euros in 2023 (multiplied by 92). For imports, the value of low-value transactions increased from 30 million euros to 735 million euros (multiplied by 24). These increases were much larger than those in total value, which multiplied by 4 for both exports and imports between 1997 and 2023. Despite these large relative increases, our second stylized fact is that low-value transactions still represent a small share of the total value of trade: 0.4\% and 0.2\% of the total value of exports and imports in 2023, respectively (panel~B).\footnote{In 1997, the figures were 0.02\% and 0.03\% for exports and imports, respectively.} \cite{fajgelbaum2024deminimis} calculated that de minimis imports represented 7.2\% of US imports of consumer goods in 2023. If we used the 800 USD threshold to define low-value transactions, these would represent 2.6\% of the Spanish imports of consumer goods in 2023. For exports, the share of low-value transactions would increase to 3.8\%. Therefore, when using the US de minimis threshold, low-value transactions represent a nonnegligible share of the Spanish trade in consumer goods.

A major limitation of our dataset is that most low-value transactions do not report the traded product. Specifically, 70\% and 71\% of low-value trade transactions in 2023 did not report the exported or imported product, respectively. This occurs because export transactions equal to or below the de minimis value of the destination country, and import transactions equal to or below 150 euros are not obliged to report the traded product. Keeping this limitation in mind, the third stylized fact is that clothing and footwear represented the largest share of low-value transactions that recorded the traded product in 2023: 41\% and 30\% of the total number of low-value export and import transactions, respectively (panel~C of Figure~\ref{fig:stylized_facts_1}). By broad economic activities, consumption goods represented 67\% and 53\% of the total number of low-value export and import transactions, respectively, in 2023; intermediate goods represented 26\% and 37\% of export and import transactions, respectively; and parts related to capital goods represented 7\% and 10\% of export and import transactions, respectively. Interestingly, in 1997, books were the top product in exports and electrical machinery in imports (panel~A of Figure~\ref{fig:stylized_facts_1997} in the Online Appendix). The share of clothing and footwear in 1997 was 12\% and 13\% for exports and imports, respectively. This comparison highlights that the clothing and footwear industries played an important role in the growth of the number of low-value transactions.\footnote{Panel~A of Figure~\ref{fig:stylized_facts_2023_high} in the Online Appendix shows that clothing and footwear also occupied prominent roles among the most exported and imported products in high-value transactions in 2023. However, their share in the total number of transactions was less than that in low-value transactions: 23\% in large-value export transactions versus 41\% in low-value transactions, and 22\% versus 30\% in import transactions.}

\begin{figure}[t!]
\begin{center}
\caption{Low-value transactions. Stylized facts II}
\label{fig:stylized_facts_2}
\hspace*{-2cm}
\begin{tabular}{c c}
\multicolumn{2}{c}{\small A. Top 20 countries by share in total number of low-value transactions (2023)}\\
\small Exports& \small Imports\\
\centering
\includegraphics[scale=.5]{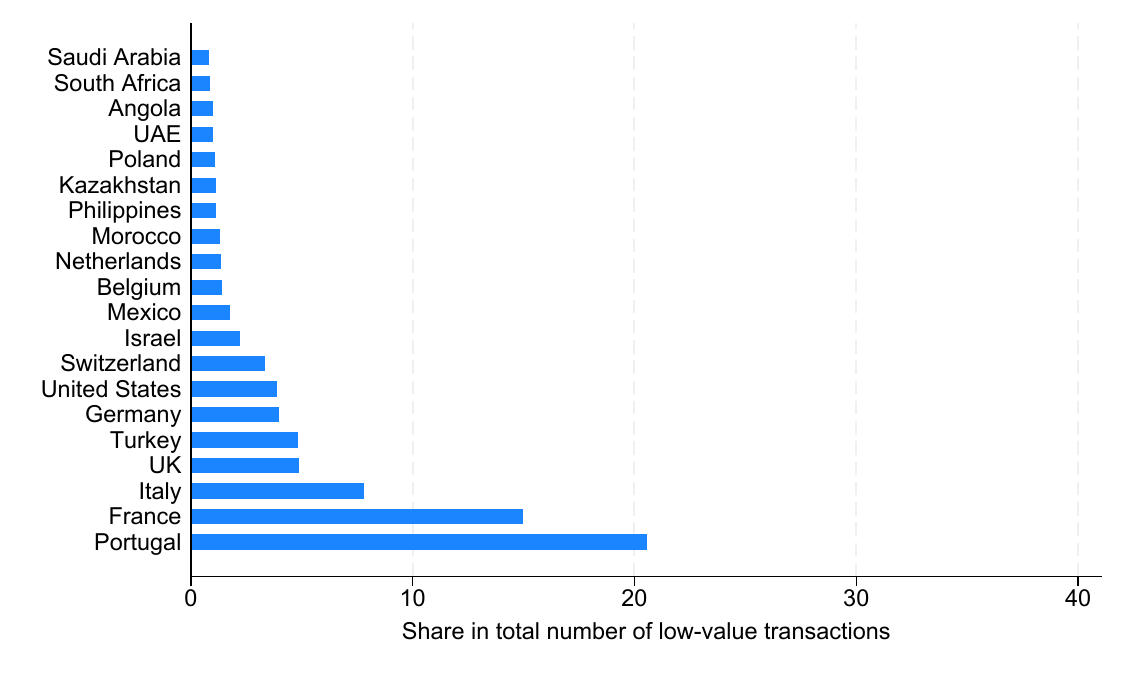}&\hspace*{-.26in}\includegraphics[scale=.5]{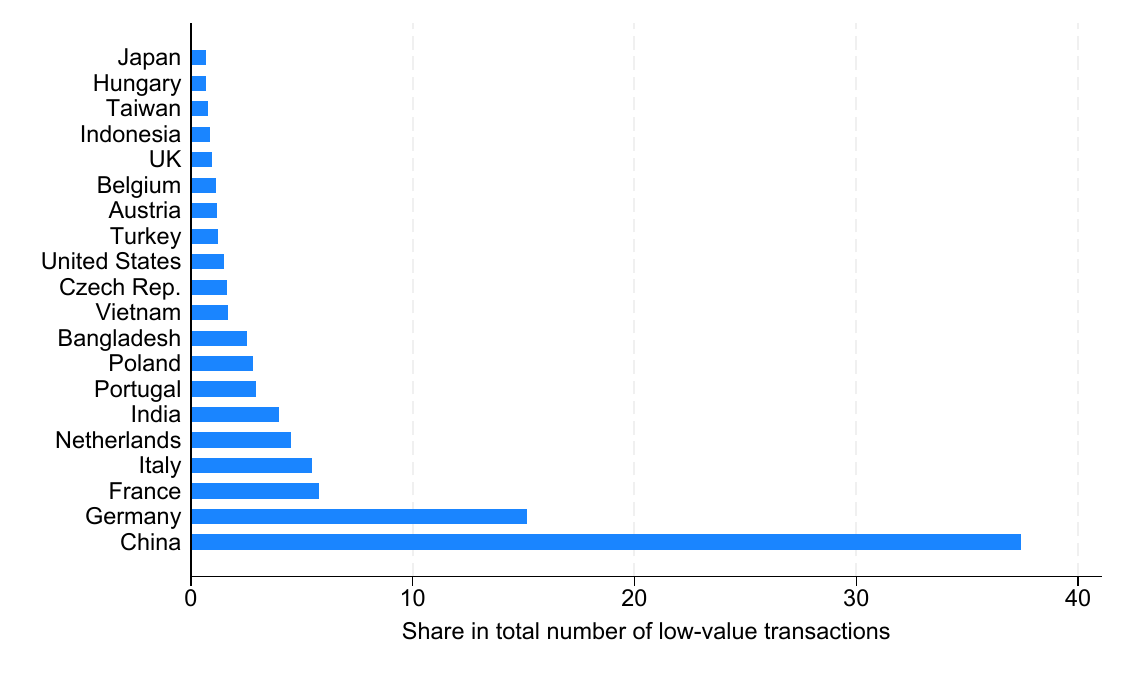}\\
\\[1em]	
\multicolumn{2}{c}{\small B. Number of firms}\\
\small Exports& \small Imports\\
\centering
\includegraphics[height=2.2in]{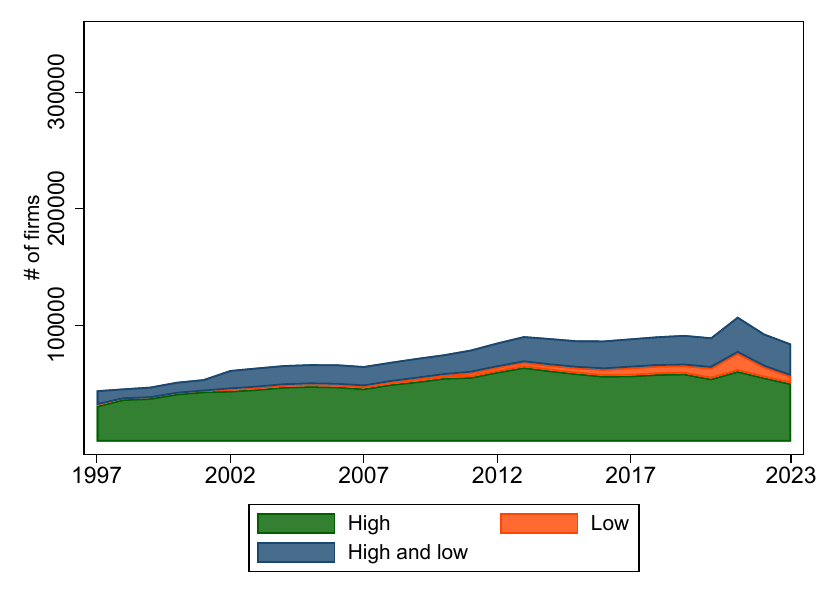}&\hspace*{-.26in}\includegraphics[height=2.2in]{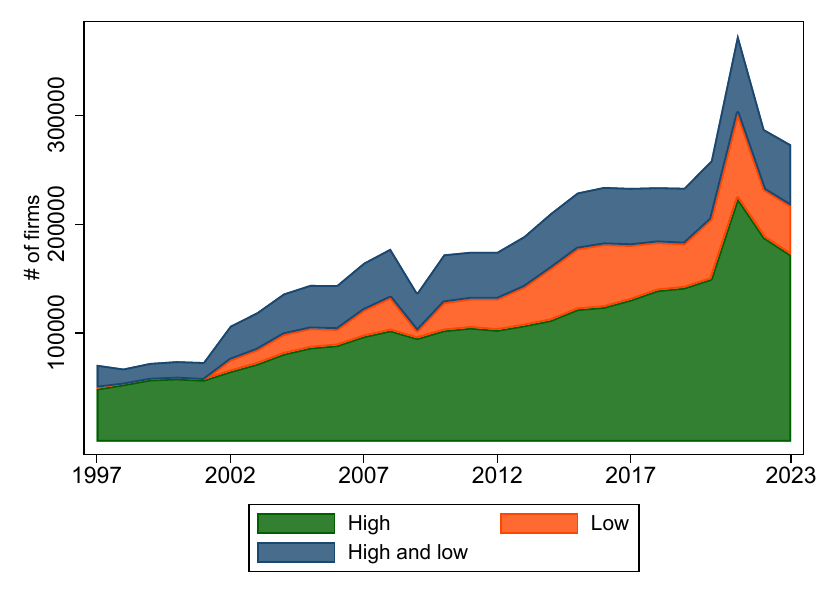}\\
\\[1em]	
\multicolumn{2}{c}{\small C. Concentration of the number of transactions by firm, 2023 (\% of the total number of transactions)}\\
\small Exports& \small Imports\\
\centering
\includegraphics[height=2.2in]{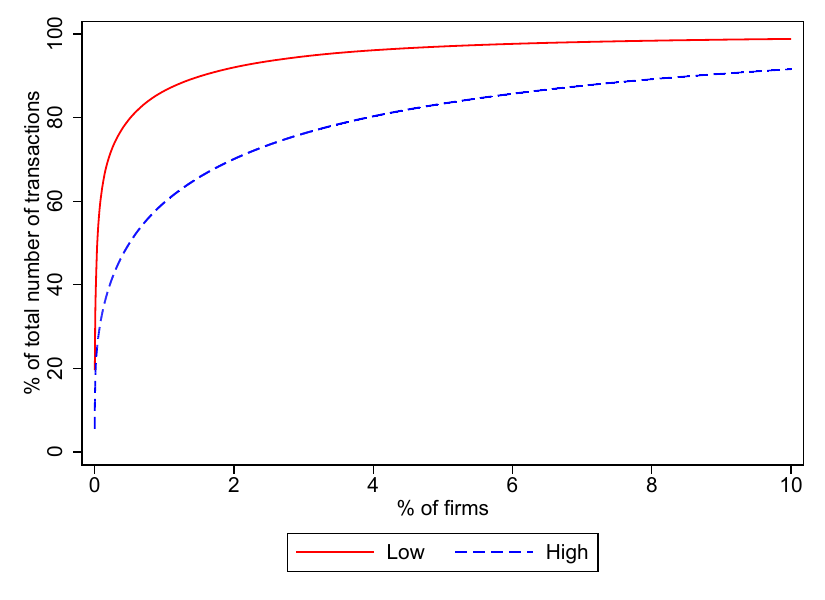}&\hspace*{-.26in}\includegraphics[height=2.2in]{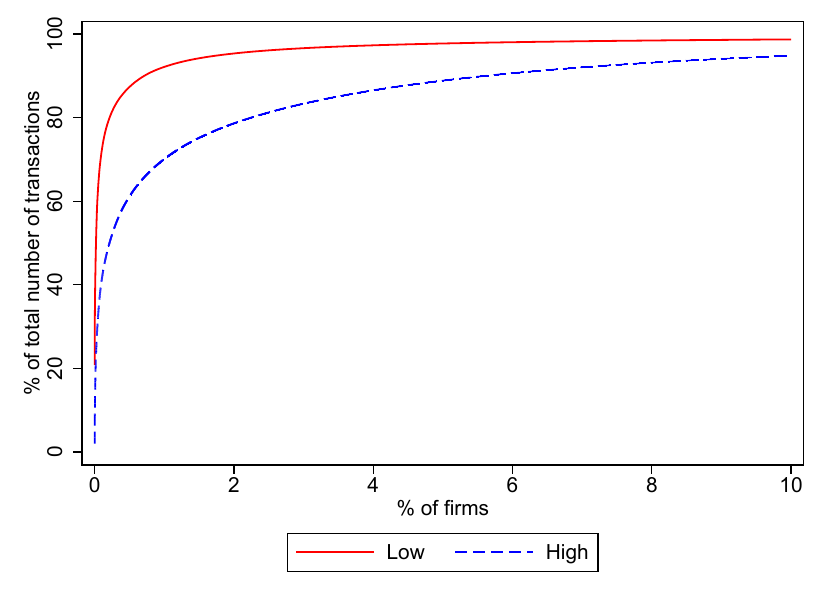}\\
\end{tabular}
\end{center}
\footnotesize Note: authors' calculations based on AEAT-Customs.
\end{figure}

The fourth stylized fact, illustrated by Panel~A of Figure~\ref{fig:stylized_facts_2}, is that Spain's neighbors, Portugal and France, were the main destinations for low-value exports in 2023 (35\% of the total number of export transactions). However, the ranking of export destinations is sensitive to the year analyzed. For example, the United States was the most important destination of low-value transactions in 2021 (Figure~\ref{fig:destinations_origins_2021} in the Online Appendix). The difference in the ranking of countries between 2021 and 2023 will be explained when discussing the causes of the increase in the number of low-value trade transactions in the following section.

China was the main source of low-value import transactions in 2023, accounting for 37\% of the total number of import transactions. Only eight countries that appear on the list of the top 20 export destinations are also included in the list of the top 20 import origins. In particular, China, which is the most important origin of low-value imports by volume, is not included in the list of the top 20 destinations.\footnote{If we added-up China and Hong-Kong, the ranking of the origins of low-value import transactions in 2021 would be similar to that in 2023 (Figure~\ref{fig:destinations_origins_2021} in the Online Appendix).} The ranking of export destinations is similar in 1997 and 2023 (panel~B of Figure~\ref{fig:stylized_facts_1997} in the Online Appendix). In contrast, China represented less than 5\% of the total number of low-value import transactions in 1997. This comparison suggests that China played a key role in the growth of the number of low-value import transactions.\footnote{Panel~B of Figure~\ref{fig:stylized_facts_2023_high} shows that China also represented a large share of high-value import transactions in 2023. However, this share was smaller than that in low-value import transactions: 20\% versus 37\%.} 

We estimated descriptive regressions to identify the country-level factors correlated with having a high share in Spain's total number of low-value export and import transactions. Table~\ref{tab:descriptive_countries} in the Online Appendix shows that the number of low-value export transactions is positively correlated with GDP and GDP per capita of the importer, sharing a land border and official language, and negatively correlated with distance and being a member of the EU. The number of low-value import transactions is positively correlated with the GDP of the source country, being a member of the EU, and negatively correlated with sharing an official language. Belonging to a free trade agreement does not play any role in either exports or imports. This is consistent with the fact that low-value import transactions are likely to be duty-free.\footnote{\cite{volpe2024consumersandfirms} get a similar result when comparing firms' and consumers' imports. The former would be equivalent to our high-value transactions and the latter to our low-value transactions.} However, we observe that belonging to the EU has a negative effect on the share of low-value export transactions in the total number of export transactions, while it has a positive effect on imports.

The fifth stylized fact, illustrated by panel~B of Figure~\ref{fig:stylized_facts_2}, is that the number of firms that only trade in low-value transactions is smaller than that that trade in high-value transactions, or in high and low-value transactions. For example, in 2023, the number of firms that only traded in low-value transactions represented 8\% and 16\% of all exporters and importers, respectively. These percentages were much higher than those at the beginning of the period (2\% for both exporters and importers). There is an increase in the percentage of exporters that trade in large- and low-value transactions. In contrast, the percentage of this category of firms decreases among importers. We find a positive correlation between the number of large- and low-value transactions performed by a firm both for exports and imports. We also find a positive correlation between the total value of large- and low-value transactions (Table~\ref{tab:reg_large_small_firm} in the Online Appendix).

Finally, the sixth stylized fact, illustrated by panel~C of Figure~\ref{fig:stylized_facts_2}, is that the concentration of the number of transactions by firm is higher in low-value transactions than in high-value ones. For example, the top firm in number of low-value export transactions represented 20\% of all low-value export transactions in 2023, while the top firm in high-value export transactions ``only'' represented 6\% of the total high-value export transactions. For imports, the differences were even larger: the top firm represented 21\% and 2\% of the total number of small- and high-value transactions, respectively. Furthermore, the concentration of transactions by firm increased between 1997 and 2023, in particular for low-value transactions: the share of the top firm in the total number of transactions increased from 3\% to 20\% in exports and from 3\% to 21\% in imports.

\section{What explains the increase in the number of low-value trade transactions?}
\label{sec:mechanism}

This section offers two explanations for the large increase in the number of low-value trade transactions between 1997 and 2023. The first addresses the rise in the number of low-value transactions in exports. As mentioned above, among the observations that reported the CN 8-digit product code, clothing and footwear represented 41\% of the total number of low-value export transactions in 2023 and explained 71\% of the increase in the number of low-value export transactions between 1997 and 2023. 

Since the beginning of the 1990s, some Spanish fashion firms have become internationally well known brands \citep{guillen2005rise,guillen2010new}. The competitiveness of these firms has been based on the so-called fast fashion strategy, which focuses on rapidly adapting the clothing displayed in the store to changing consumer preferences \citep{crofton2007zara}. In addition, these firms generate a sense of scarcity with collections that end quickly, creating a fear of missing out feeling among customers, which leads them to purchase a garment without delay in the event that it is no longer available on the next visit to the store \citep{ghemawat2006zara}.\footnote{As explained by \cite{fernandes2020fastfashion}, the fast fashion strategy may also have helped garment manufacturers in developed countries respond to low-cost competition from Asia.} 

The fast-fashion business model requires stores to replenish frequently, a procedure that echoes the just-in-time system, introduced first by Toyota in the automotive sector and later implemented in other industries \citep{womack1990machine}. Since all products exported by leading Spanish retailers go through a distribution center in Spain \citep{ghemawat2006zara,rodriguez2024mango}, we expect the fast-fashion strategy to generate a large number of low-value export transactions between the distribution center and stores abroad. We also anticipate the frequency of transactions to be higher in destinations geographically far from Spain. Because products take longer to reach their destination, firms have to respond very quickly to replenishment demands from stores to ensure that preferences do not change before goods arrive. This will require the use of air transport to serve destinations that cannot be reached by road in a few days from Spain. 

\begin{figure}[t!]
\begin{center}
\caption{Mechanisms explaining the increase in the number of low-value transactions}
\label{fig:mechanisms}
\hspace*{-2cm}
\begin{tabular}{c}
\small A. Number of low-value export transactions from the leading fashion retailers and from other firms, 1997-2023\\
\includegraphics[height=2.3in]{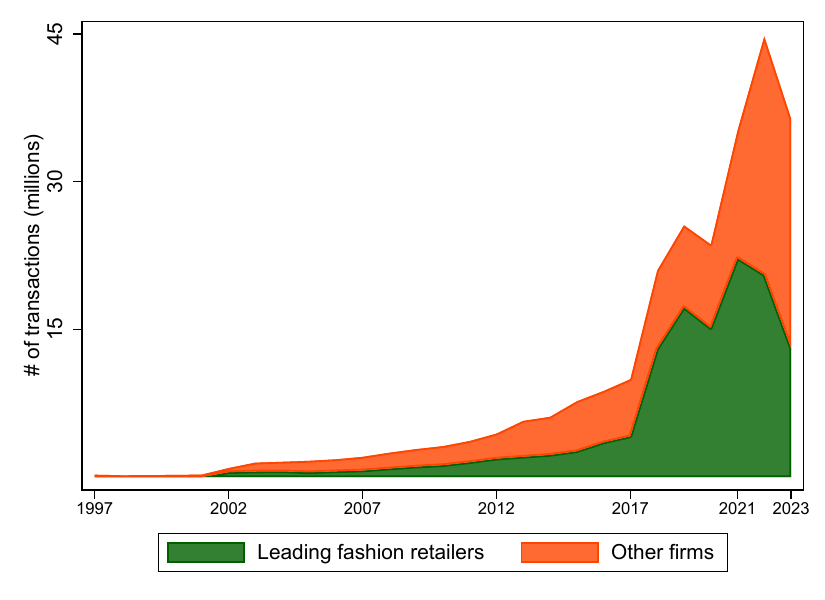}\\
\\[.5em]	
\small B. Top 10 destinations by volume of low-value export transactions from leading fashion retailers, 2021 (millions)\\
\includegraphics[height=2.5in]{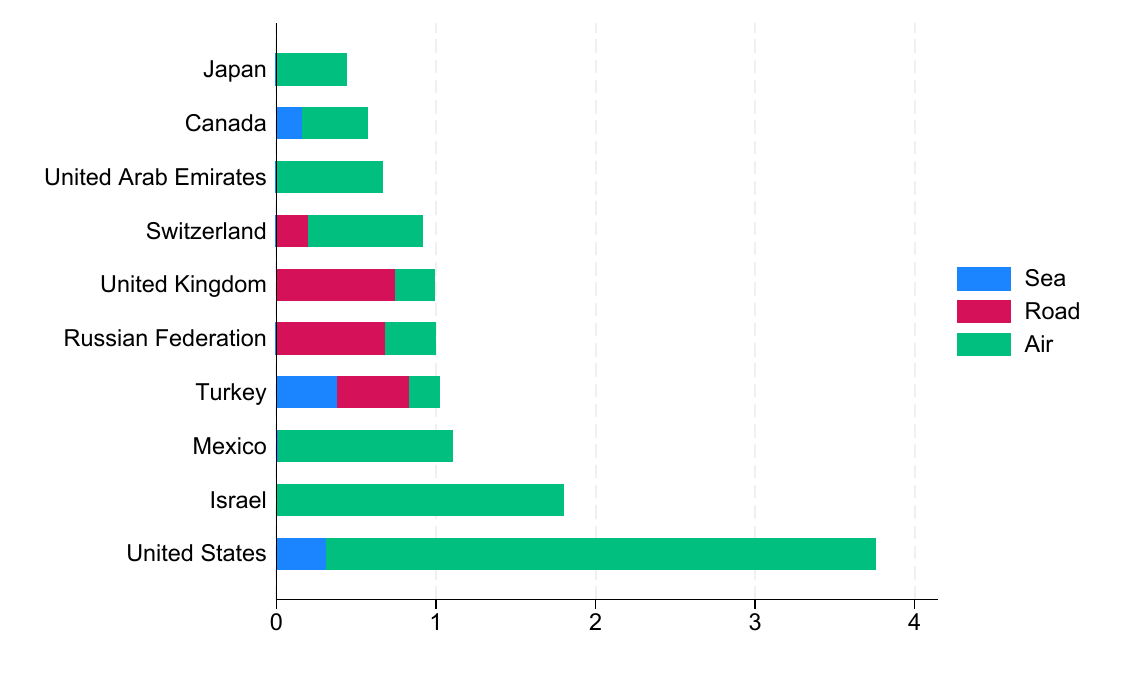}\\
\\[.5em]	
\small C. Number and share of low-value transactions in Spain's imports from China, 1997-2023\\
\includegraphics[height=2.5in]{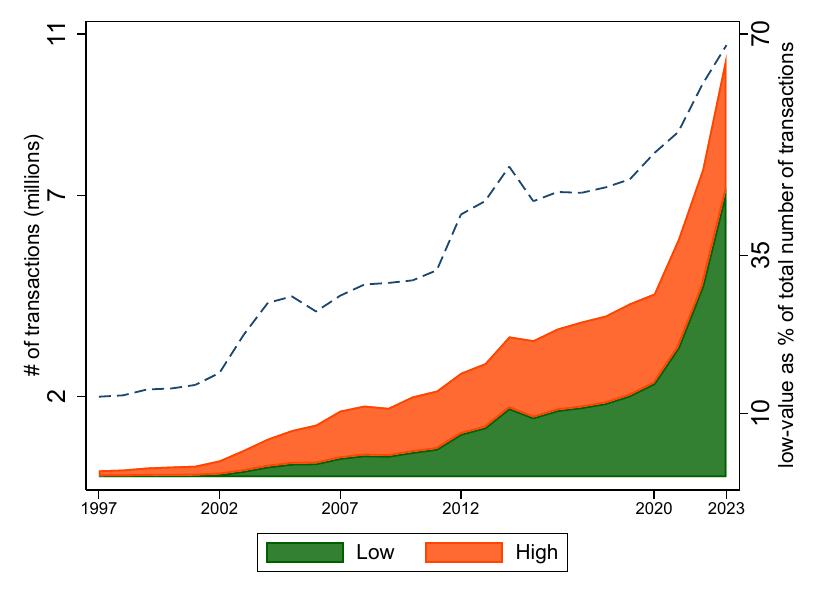}\\
\end{tabular}
\end{center}
\footnotesize Note: authors' calculations based on AEAT-Customs.
\end{figure}

Using the procedure described in \cite{delucio2018prices}, we identified the export transactions of the leading fashion retailers in Spain in our database. Panel~A of Figure~\ref{fig:mechanisms} shows the evolution of the number of low-value export transactions of the leading fashion retailers and all firms between 1997 and 2023. From 1997 to 2021, the increase in the number of low-value export transactions for all firms runs parallel to the increase in the number of low-value export transactions from the leading fashion retailers. This is because exports from the leading fashion retailers represented most of the volume of low-value export transactions: 63\% in 2021. Panel~B of Figure~\ref{fig:mechanisms} shows that the top 10 destinations by volume of low-value export transactions of the leading fashion retailers in 2021 were countries that were not geographically close to Spain.\footnote{The top 10 destinations accounted for 55\% of the total number of low-value export transactions.} In fact, countries that are closer to Spain, such as France, Germany, Italy, or Portugal, and which were some of the most important export destinations by value of the leading fashion retailers in 2021, were not included in the top 10 ranking of low-value export transaction destinations. Furthermore, except for destinations that could be reached by road in a few days from Spain (Russian Federation, Turkey, and the UK), most of the products were transported by air. These findings are in line with the fast-fashion strategy described above. In summary, from 1997 to 2021, the leading fashion retailers were responsible for the increase in the total number of low-value export transactions. The frequency of low-value transactions, their geographical distribution, and the means used to transport the products are consistent with the fast-fashion strategy followed by these firms.


The number of low-value export transactions of the leading Spanish retailers decreases from 2021 onward. A factor contributing to this decline was the termination or reduction of operations in Russia after the invasion of Ukraine in February 2022 \citep{delucio2024russia}.\footnote{Export transactions from the leading fashion retailers to Russia decreased in one million after the invasion of Ukraine.} Furthermore, the total number of low-value export transactions decreases less than the number of low-value export transactions of the leading fashion retailers. This indicates that firms not related to the fashion industry began to account for a larger share of low-value export transactions. To explain this fact, we need to introduce the second explanation for the increase in the number of low-value transactions. We do it analyzing the evolution of low-value transactions in imports.

As mentioned in the previous section, China was the most important origin, by number, of Spain's low-value import transactions in 2023. Furthermore, China was the largest contributor to the increase in the number of low-value import transactions between 1997 and 2023 (38\%). Panel~C of Figure~\ref{fig:mechanisms} shows the evolution in the number of low- and high-value import transactions from China, and the share of low-value transactions in the total number of import transactions from China. In 1997, low-value import transactions only represented 12\% of the total number of import transactions from China. By 2023, they represented 68\% of the total number of import transactions. There is a steady increase in the number of low-value import transactions from China until 2020. However, in the following three years, the number of low-value import transactions multiplied by more than three, from 2 million to 7 million, and the share of low-value import transactions in the total number of import transactions increased by 17 percentage points. 

This staggering increase coincides with the popularization of online retail platforms founded in China, such as Shein or Temu.\footnote{For example, in August 2023, Temu and Shein were the first and third most downloaded applications in Spain (see "Temu, Shein, Miravia: el agresivo comercio electrónico chino que rivaliza con Amazon" by Pablo G. Bejarano, www.elpais.com, 11 November 2023 at \url{https://elpais.com/tecnologia/2023-11-11/temu-shein-miravia-el-agresivo-comercio-electronico-chino-que-rivaliza-con-amazon.html}.} These online platforms allow consumers to access a wide range of low-value Chinese products while providing Chinese manufacturers with access to a global market. They follow a strategy that resembles an extreme version of the fast-fashion strategy, introducing thousands of new products every day and offering ``cannot-miss'' promotions and discounts. Transactions mediated through these platforms are denoted as direct to the consumer because purchases go directly from the overseas producer to the local consumer. 


Our data set does not provide information on whether a transaction is the result of an online or an offline purchase. Having this limitation in mind, we examine whether other facts are consistent with the direct-to-consumer explanation for the rise in the number of low-value import transactions from China from 2020 onward. In 2023, a firm concentrated almost half of low-value import transactions originating in China. This firm did not receive imports from China in a specific province in Spain, but they were scattered throughout all Spanish provinces. Furthermore, this firm recorded 300,000 different import transactions from China in each monthly customs declaration. This very large number of declared transactions is consistent with the fact that online platforms, or their logistic partners, can fulfill the customs declaration on behalf of their final customers, when the transaction is equal or below the EU de minimis threshold. It is very unlikely that a firm would fraction imports for its own use into so many low-value transactions to avoid the payment of tariffs, because such a large number of transactions would raise the suspicion of tax authorities. Therefore, the geographical distribution and the large number of transactions are consistent with the argument that products were sent directly from Chinese producers to Spanish customers and that the importer was an operator that facilitated customs clearance and shipments.

The top importer by number of low-value transactions from China in 2023 was the same as in 2022. However, before 2022, the main importers were a large retailer and two fashion firms. These firms received all low-value imports from China in the province where their headquarters and main warehouse were located. We do not have information to determine whether the increase in the number of low-value import transactions from China of these firms was explained by the expansion of their E-commerce sales or the use of a just-in-time strategy. However, there are two facts that are consistent with the argument that these purchases were likely to be linked to these firms' E-commerce activities. First, in the case of the large retailer, the number of low-value import transactions from China was almost negligible in 2019. However, it became the importer with the largest number of low-value transactions from China in 2020. That year, due to Covid, non-essential stores remained closed for some weeks. Furthermore, even when the stores opened after the strictest confinement measures were lifted, the shopping experience became less pleasant for the customers due to health-related restrictions. In parallel, the retailer reported a very large increase in E-commerce sales between 2019 and 2020. These facts suggest that it was the increase in purchases on this firm's online marketplace that contributed to the increase in the number of low-value import transactions from China. Second, as argued above, the fact that the large retailer and the two fashion firms reported very large numbers of low-value import transactions in their customs declarations is consistent with the documentation of low-value E-commerce-based transactions.


We find that the four firms that explain the rise in the number of low-value import transactions from China are also responsible for one third of the total increase in the number of low-value import transactions between 1997 and 2023. The remaining 100,000 firms that performed low-value import transactions each contributed a very tiny fraction to the remaining two-thirds of the increase.\footnote{The exceptions are two firms that represented 4\% and 2\% of the increase and specialize in low-value imports from Germany.}

At this point, we address the increase in the number of low-value export transactions that occurred between 2021 and 2023. The firm that contributes the most to the growth of low-value transactions in this period (16\% of the total) is a retailer. The increase in the number of low-value export transactions coincides with the opening of online marketplaces in three European countries by this firm since 2021. These marketplaces allowed Spanish firms that supplied the retailer in Spain to sell their products on other European markets. Consistent with this narrative, we observe a large increase in the number of low-value export transactions by the retailer from 2021 onward, particularly in those European countries in which a marketplace was launched. The emergence of this retailer as a major player in the volume of export transactions also explains why there are differences in the ranking of destinations for low-value export transactions between 2021 and 2023. In 2021, low-value export transaction destinations were still dominated by the destinations of leading fashion retailers (panel~B of Figure~\ref{fig:mechanisms}). However, by 2023, export destinations were also affected by the new retailer's destinations, which were concentrated in European countries geographically close to Spain.

Taking into account the caveats of our data set, which cannot discriminate between online and offline transactions, these results suggest that online marketplaces could also play a relevant role in explaining the increase in the number of low-value transactions in exports. They also underscore how decisions taken by few firms can have very large effects on the number of low-value trade transactions.

\section{Conclusion}
\label{sec:conclusion}
This paper has documented a novel feature of international trade: the increase in the share of low-value transactions in the volume of transactions. Using the universe of Spanish international trade transactions, we have shown that the share of low-value transactions in the total number of transactions increased from 9\% to 61\% in exports and from 14\% to 54\% in imports between 1997 and 2023. 


We have offered two explanations for the increase in the number of low-value trade transactions. First, online retail platforms have allowed consumers to access a wide range of low-value items manufactured abroad, triggering a large increase in the volume of low-value transactions. Second, the fast-fashion strategy adopted by clothing retailers, which uses a frequent stocking of stores to adjust for changes in consumer preferences, has also led to a large increase in low-value transactions.  

The increase in the share of low-value transactions can have important implications on international trade. Online platforms have allowed consumers, particularly those with low income, to have a more direct involvement in trade, providing them with a greater leverage in shaping trade flows. In addition, the increase in low-value import transactions may lead countries to adopt protection measures by reducing the de minimis threshold. Finally, it has also raised concerns about the environmental impact of trade and data protection.   



\bibliography{citations}

\clearpage
\onehalfspacing

\begin{center}
\textbf{\Large \centering Online Appendix for ``The increasing share of low-value transactions in international trade''}
\end{center}

This online appendix displays additional Tables and Figures mentioned in the main text.

\renewcommand\thesection{\Alph{section}}

\section*{Additional tables}
\label{app:additional_tables}

\setcounter{figure}{0}
\renewcommand\thefigure{A.\arabic{figure}}

\setcounter{table}{0}
\renewcommand\thetable{A.\arabic{table}}

\begin{table}[htbp]
\begin{center}
\footnotesize
\caption{Estimates from descriptive regressions analyzing the factors affecting a partner's share in the total number of low-value transactions}
\label{tab:descriptive_countries}
{
\def\sym#1{\ifmmode^{#1}\else\(^{#1}\)\fi}
\begin{tabular}{l*{2}{c}}
\hline\hline
                    &\multicolumn{1}{c}{(1)}&\multicolumn{1}{c}{(2)}\\
                    &\multicolumn{1}{c}{Exports}&\multicolumn{1}{c}{Imports}\\
\hline
GDP (ln)            &       0.548\sym{a}&       0.858\sym{a}\\
                    &     (0.057)       &     (0.283)       \\
[1em]
GDP pc (ln)         &       0.400\sym{a}&      -0.113       \\
                    &     (0.093)       &     (0.451)       \\
[1em]
Distance (ln)       &      -0.655\sym{a}&       0.370       \\
                    &     (0.125)       &     (0.426)       \\
[1em]
Contiguity          &       1.362\sym{b}&       0.806       \\
                    &     (0.565)       &     (0.615)       \\
[1em]
Language            &       1.007\sym{a}&      -2.064\sym{a}\\
                    &     (0.284)       &     (0.591)       \\
[1em]
European Union      &      -1.474\sym{a}&       2.276\sym{a}\\
                    &     (0.274)       &     (0.595)       \\
[1em]
Free trade agreement&       0.019       &      -0.607       \\
                    &     (0.239)       &     (0.912)       \\
\hline
Observations        &         192       &         185       \\
Pseudo-R2           &       0.668       &       0.597       \\
\hline\hline
\end{tabular}
}

\caption*{\begin{footnotesize}Note: Data is for 2021. In columns~1 and~2 the dependent variable is the share of a country in Spanish total number of low-value export and import transactions, respectively. We estimate a Poisson pseudo-maximum likelihood model. Distance is the harmonic mean of the population-weighted distance between the most populated cities. Contiguity is an indicator variable that turns one if the partner country shares a land border with Spain. Language is an indicator variable that turns one if Spain and the partner country share the same official language. European Union is an indicator variable that turns one if the partner country belongs to the European Union. Free trade agreement is an indicator variable that turns one if Spain and the partner country belong to the same free trade agreement. GDP and GDP per capita data are from \href{https://databank.worldbank.org/source/world-development-indicators}{World Bank's World Development Indicators} database. The remaining independent variables are from the \href{http://www.cepii.fr/CEPII/en/bdd_modele/bdd_modele.asp}{CEPII gravity} database. Standard errors are in parentheses. a, b, and c: statistically significant at 1\%, 5\%, and 10\%, respectively.\end{footnotesize}}
\end{center}
\end{table}

\begin{table}[htbp]
\begin{center}
\footnotesize
\caption{Correlation between large- and low-value transactions at the firm level}
\label{tab:reg_large_small_firm}
{
\def\sym#1{\ifmmode^{#1}\else\(^{#1}\)\fi}
\begin{tabular}{l*{4}{c}}
\hline\hline
                    &\multicolumn{2}{c}{Number}             &\multicolumn{2}{c}{Value}              \\\cmidrule(lr){2-3}\cmidrule(lr){4-5}
                    &\multicolumn{1}{c}{(1)}&\multicolumn{1}{c}{(2)}&\multicolumn{1}{c}{(3)}&\multicolumn{1}{c}{(4)}\\
                    &\multicolumn{1}{c}{Exports}&\multicolumn{1}{c}{Imports}&\multicolumn{1}{c}{Exports}&\multicolumn{1}{c}{Imports}\\
\hline
Large value$_{ft}$  &     0.43080\sym{a}&    19.69778\sym{a}&     0.00016\sym{a}&     0.00060\sym{a}\\
                    &   (0.06503)       &   (1.84782)       &   (0.00005)       &   (0.00019)       \\
\hline
Observations        &     1504377       &     2760297       &     1504377       &     2760297       \\
Pseudo-R2           &     0.72290       &     0.54298       &     0.83725       &     0.67368       \\
\hline\hline
\end{tabular}
}

\caption*{\begin{footnotesize}Note: In columns~1 and~2 the dependent variable is the number of low-value export and import transactions in millions per firm$\times$year, respectively. In columns~3 and~4 the dependent variable is the total value in million euros of low-value export and import transactions per firm$\times$year, respectively. We estimate a Poisson pseudo-maximum likelihood model. All regressions include firm and year fixed effects. Standard errors are in parentheses. a, b, and c: statistically significant at 1\%, 5\%, and 10\%, respectively.\end{footnotesize}}
\end{center}
\end{table}

\clearpage

\section*{Additional figures}
\label{app:additional_figures}

\setcounter{figure}{0}
\renewcommand\thefigure{A.\arabic{figure}}

\setcounter{table}{0}
\renewcommand\thetable{A.\arabic{table}}

\begin{figure}[h!]
\begin{center}
\caption{Robustness analyses on the share of low-value transactions in the total number of trade transactions}
\label{fig:appendix_rob_deminimis}
\begin{tabular}{c c}
\multicolumn{2}{c}{A. Alternative low-value transaction thresholds}\\
Exports&Imports\\
\includegraphics[height=2.2in]{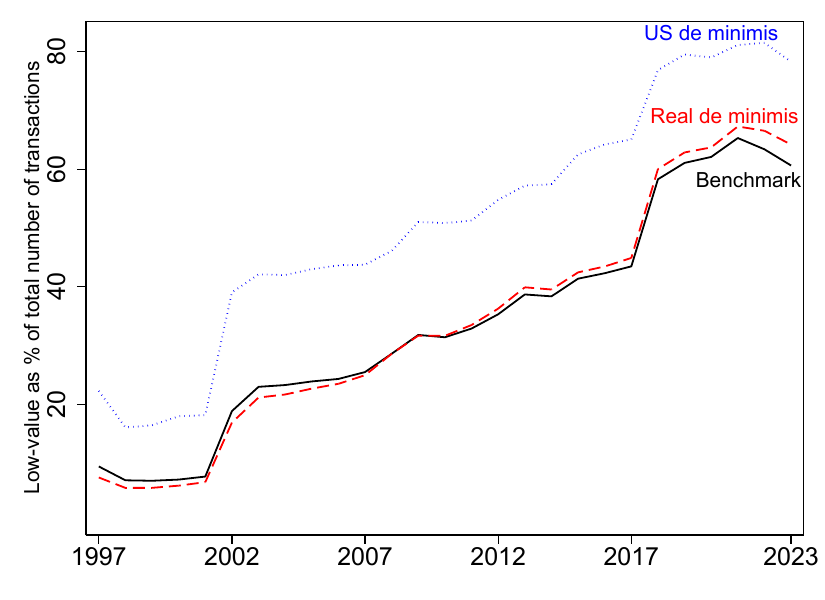}&
\includegraphics[height=2.2in]{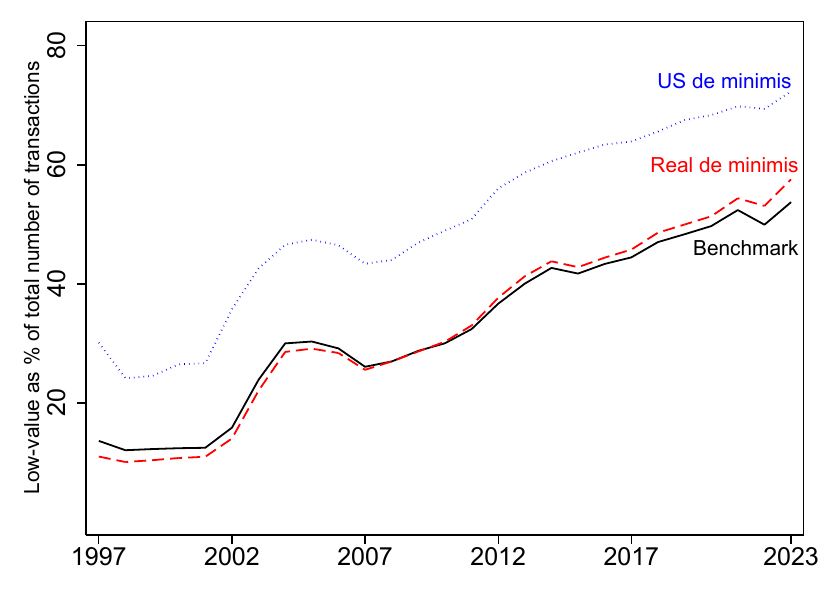}\\
\\[1em]	
\multicolumn{2}{c}{B. Alternative samples: EU27 vs. non-EU27}\\
Exports&Imports\\
\includegraphics[height=2.2in]{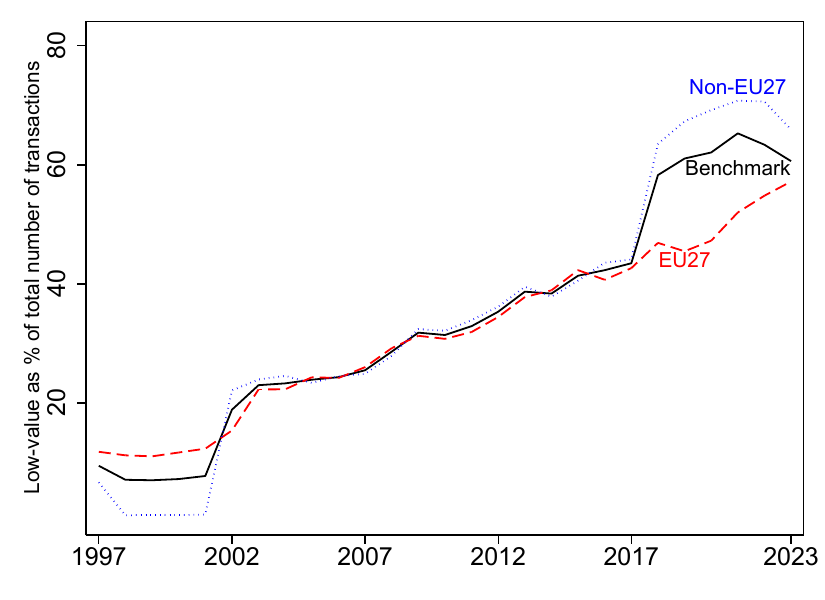}&
\includegraphics[height=2.2in]{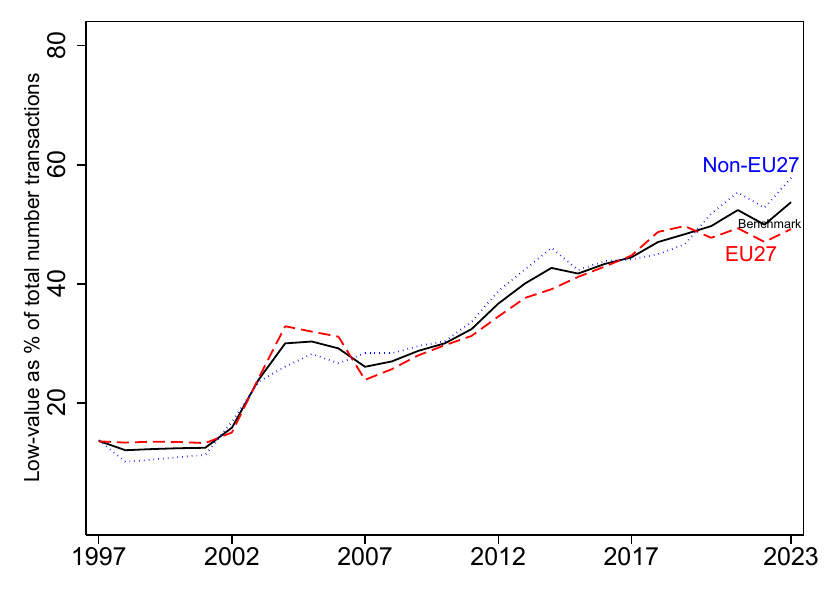}\\

\end{tabular}
\end{center}
\footnotesize Note: Real de minimis corresponds to the 150 euro threshold, established in 2009, adjusted by the Spanish consumer price index (Spanish Statistical Institute, available at \url{https://www.ine.es}). US de minimis is the 800 USD de minimis value established by the United States in 2016. Benchmark corresponds to the share of low-value transactions reported in Panel~A of Figure~\ref{fig:stylized_facts_1} in the main text.
\end{figure}

\begin{figure}[t!]
\begin{center}
\caption{Stylized facts in 1997}
\label{fig:stylized_facts_1997}
\hspace*{-2cm}
\begin{tabular}{c c}
\multicolumn{2}{c}{\small A. Top 20 HS 2-digit chapters by share in total number of low-value transactions}\\
\small Exports& \small Imports\\
\centering
\includegraphics[scale=.5]{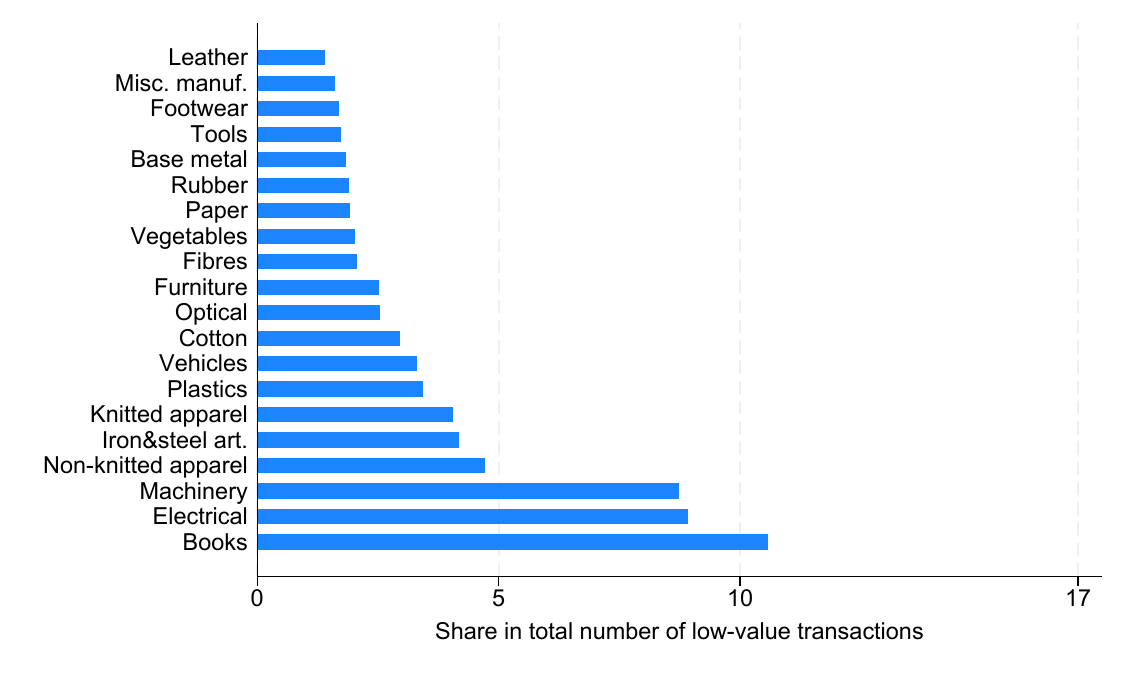}&\hspace*{-.26in}\includegraphics[scale=.5]{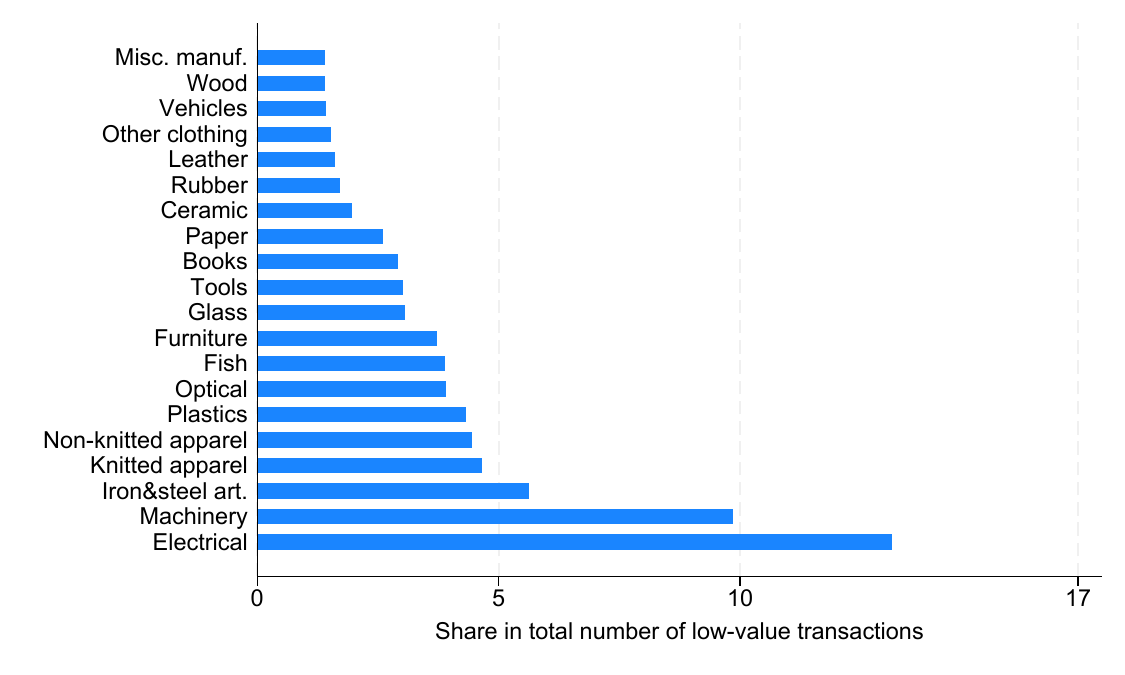}\\
\\[1em]	
\multicolumn{2}{c}{\small B. Top 20 countries by share in the total number of low-value transactions}\\
\small Exports& \small Imports\\
\centering
\includegraphics[scale=.5]{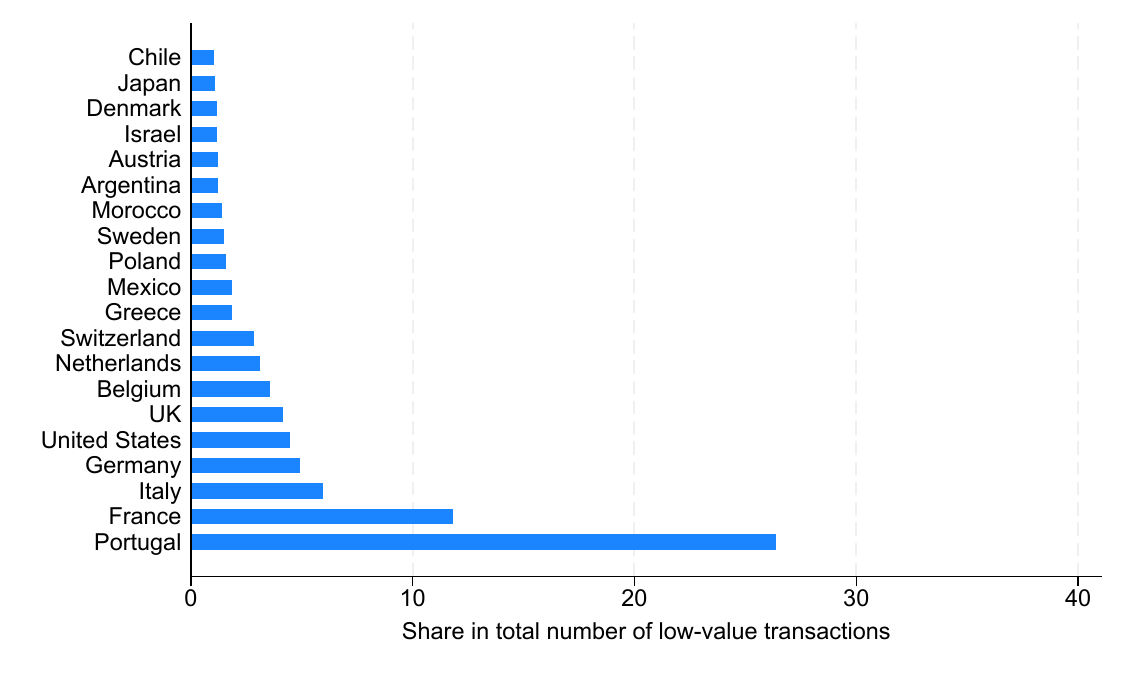}&\hspace*{-.26in}\includegraphics[scale=.5]{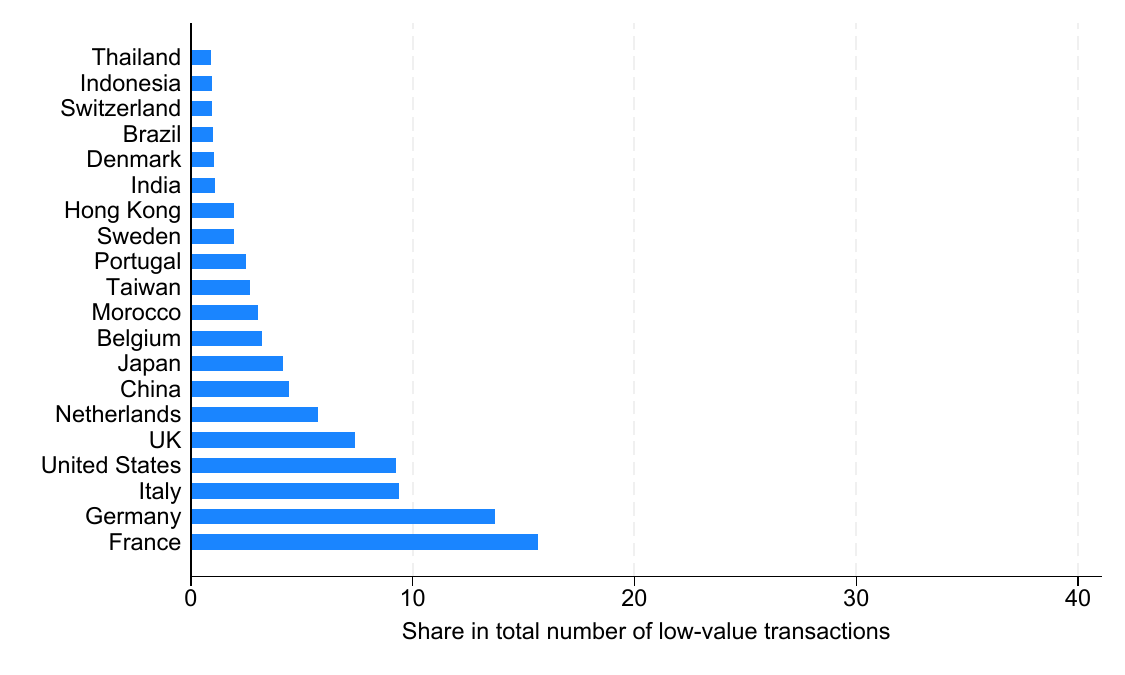}\\
\\[1em]	
\multicolumn{2}{c}{\small C. Concentration of the number of transactions by firm (\% of total number of transactions)}\\
\small Exports& \small Imports\\
\centering
\includegraphics[height=2.2in]{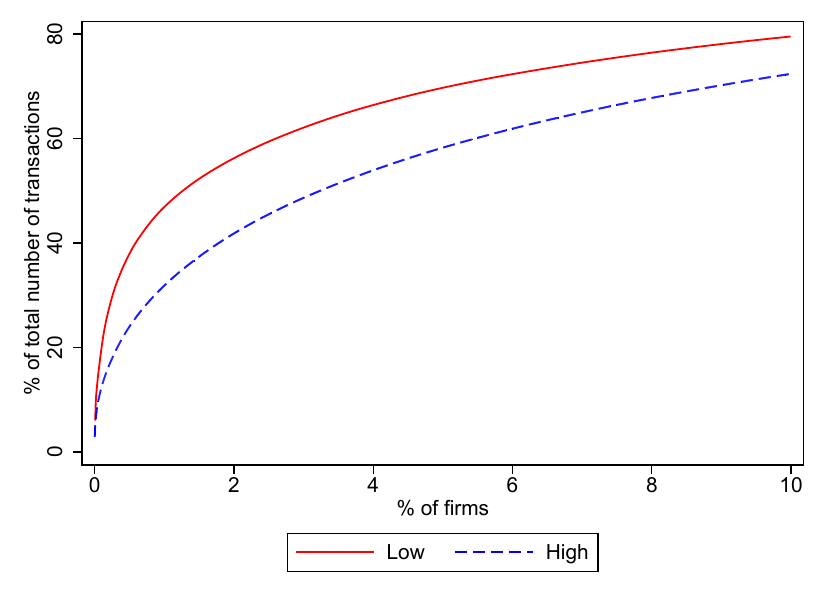}&\hspace*{-.26in}\includegraphics[height=2.2in]{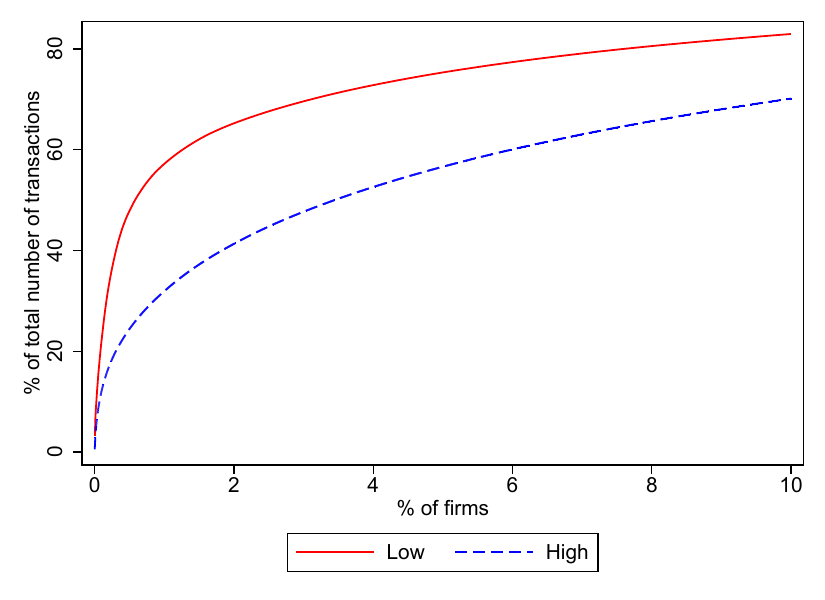}\\
\end{tabular}
\end{center}
\footnotesize Note: authors' calculations based on AEAT-Customs.
\end{figure}

\begin{figure}[t!]
\begin{center}
\caption{Stylized facts on high-value transactions in 2023}
\label{fig:stylized_facts_2023_high}
\hspace*{-2cm}
\begin{tabular}{c c}
\multicolumn{2}{c}{\small A. Top 20 HS 2-digit chapters by share in total number of high-value transactions}\\
\small Exports& \small Imports\\
\centering
\includegraphics[scale=.5]{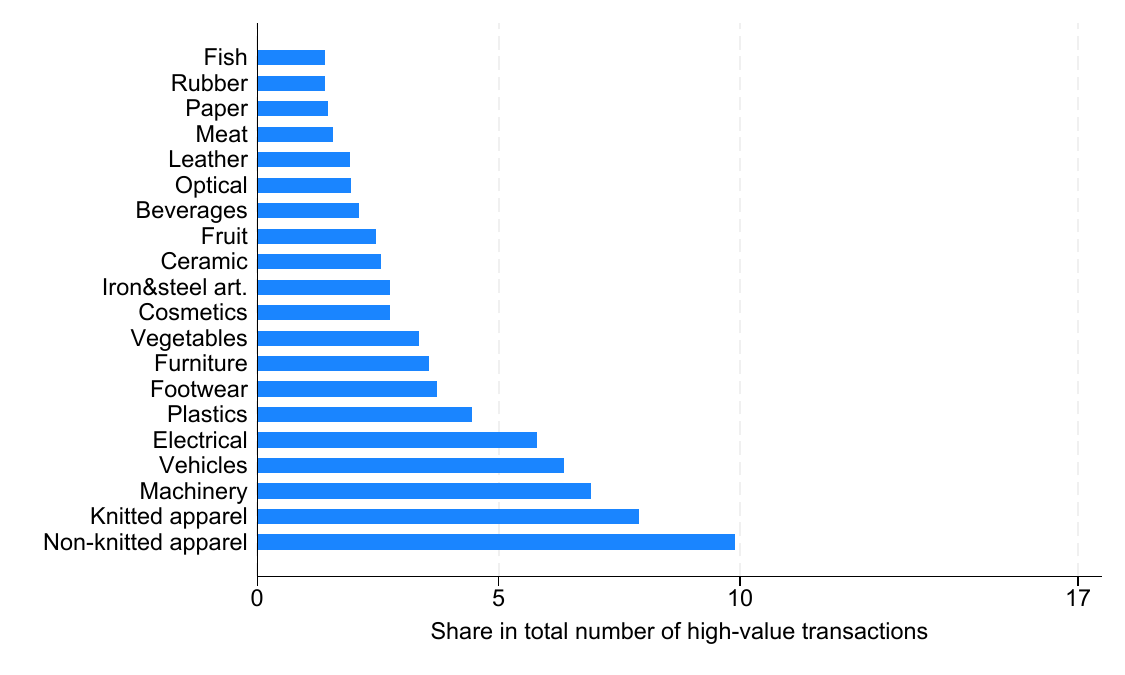}&\hspace*{-.26in}\includegraphics[scale=.5]{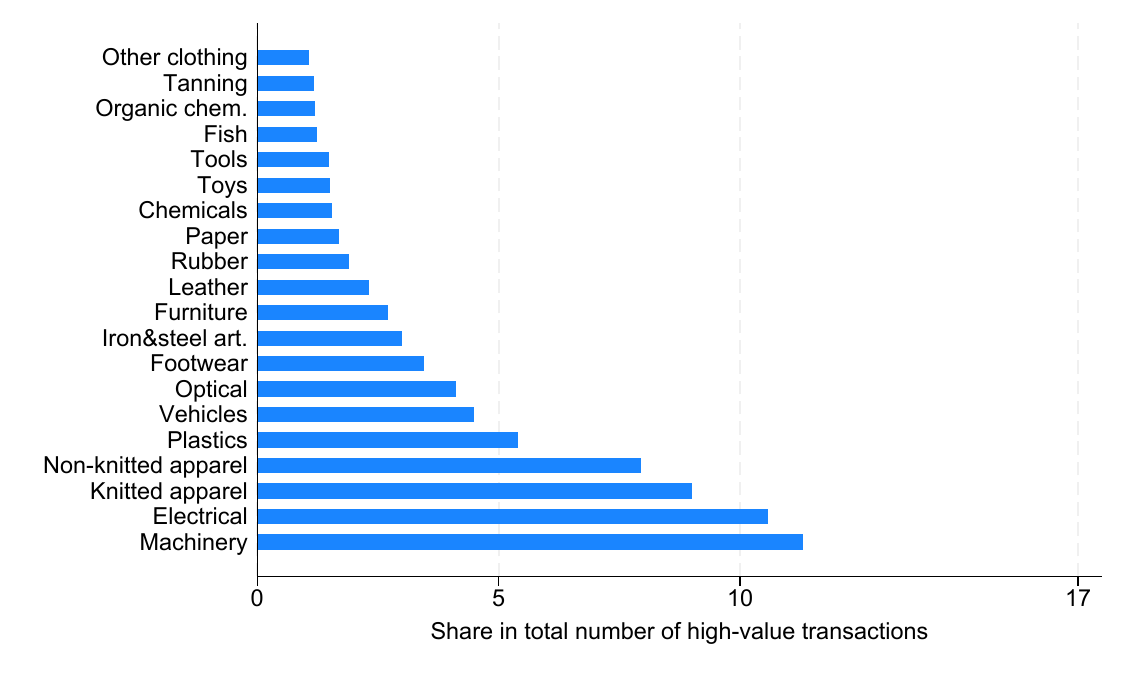}\\
\\[1em]	
\multicolumn{2}{c}{\small B. Top 20 countries by share in the total number of high-value transactions}\\
\small Exports& \small Imports\\
\centering
\includegraphics[scale=.5]{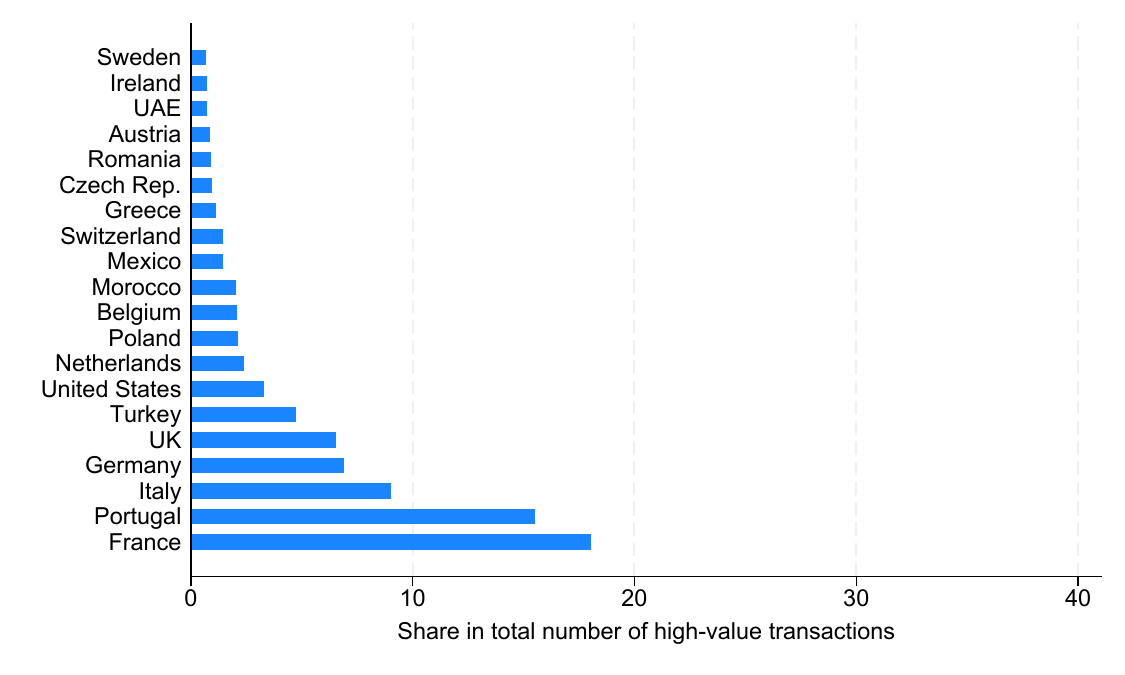}&\hspace*{-.26in}\includegraphics[scale=.5]{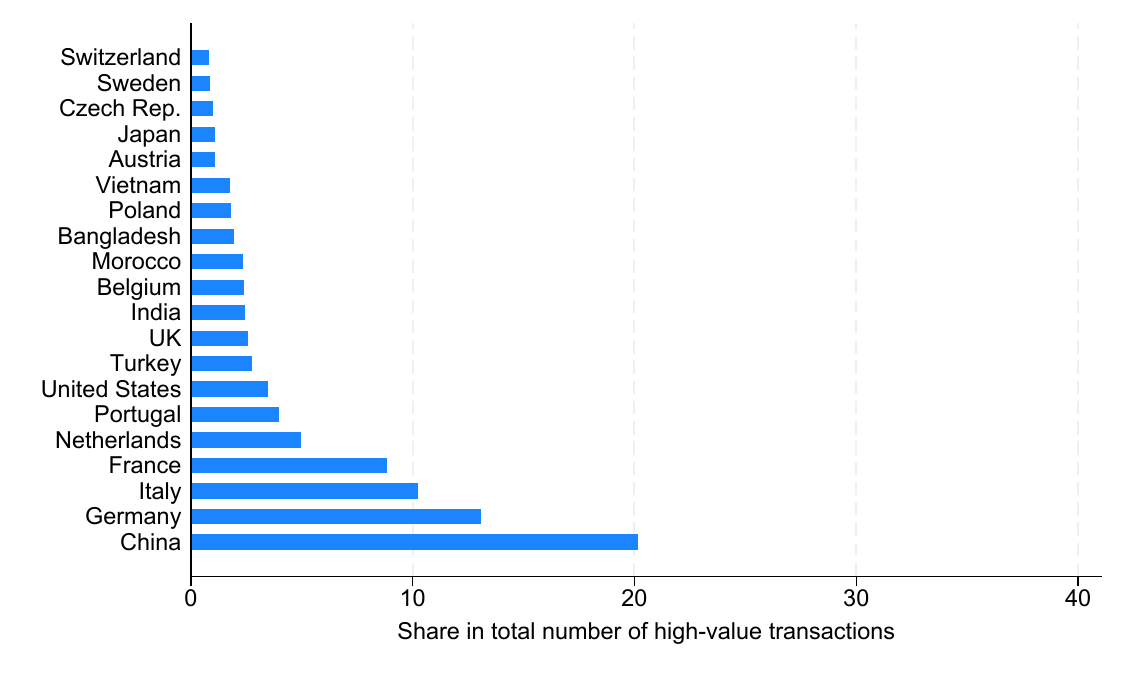}\\
\end{tabular}
\end{center}
\footnotesize Note: authors' calculations based on AEAT-Customs.
\end{figure}

\begin{figure}[t!]
\begin{center}
\caption{Ranking of destinations and origins by number of low-value transactions, 2021}
\label{fig:destinations_origins_2021}
\hspace*{-2cm}
\begin{tabular}{c c}
\small Exports& \small Imports\\
\centering
\includegraphics[scale=.5]{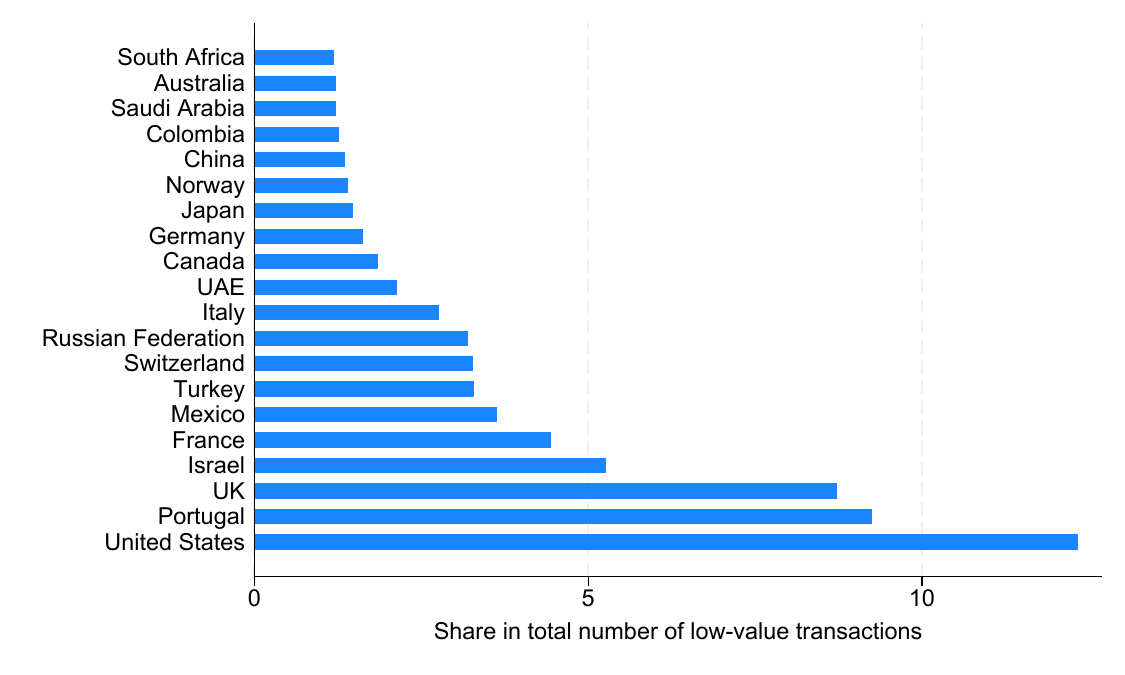}&\hspace*{-.26in}\includegraphics[scale=.5]{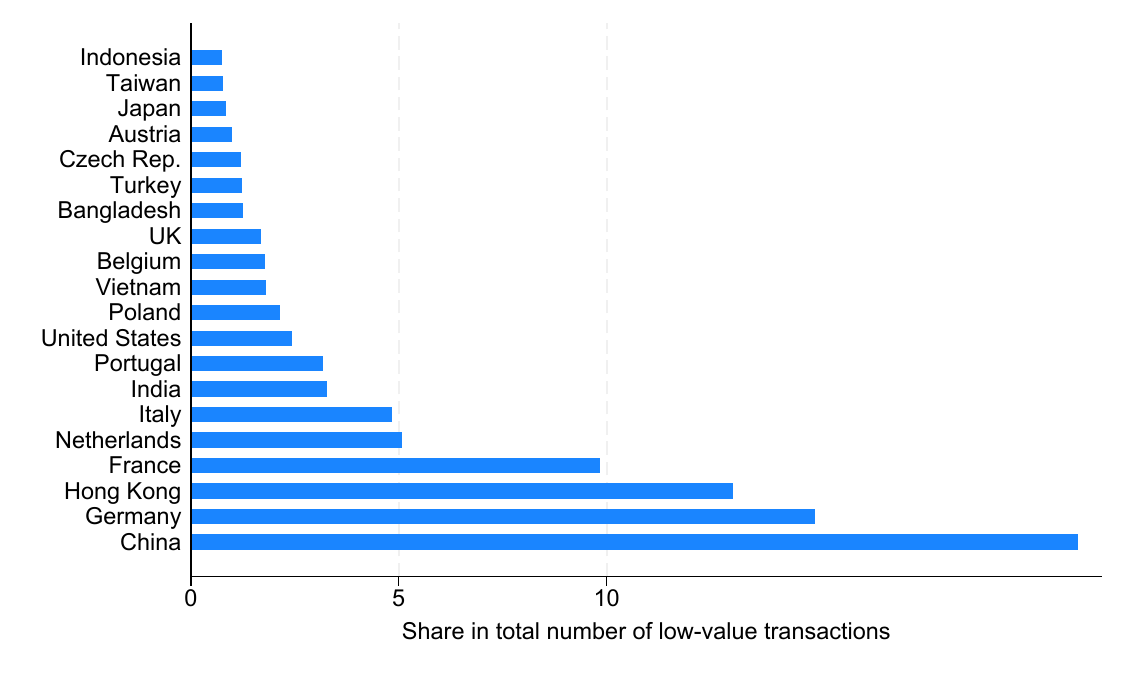}\\
\end{tabular}
\end{center}
\footnotesize Note: authors' calculations based on AEAT-Customs.
\end{figure}

\end{document}